\documentclass[titlepage,12pt]{article}
\usepackage{amsfonts}
\usepackage{geometry}
\usepackage[onehalfspacing]{setspace}
\usepackage{times}
\usepackage{mathrsfs}

\newtheorem{theorem}{Theorem}

\input{tcilatex}
\renewcommand{\mathcal}{\mathscr}

\begin{document}

\title{Warped Functional Regression}
\author{Daniel Gervini \\
Department of Mathematical Sciences,\\
University of Wisconsin--Milwaukee,\\
PO Box 413, Milwaukee, Wisconsin 53201, USA\\
gervini@uwm.edu}
\maketitle

\begin{abstract}
A characteristic feature of functional data is the presence of phase
variability in addition to amplitude variability. Existing functional
regression methods do not handle time variability in an explicit and
efficient way. In this paper we introduce a functional regression method
that incorporates time warping as an intrinsic part of the model. The method
achieves good predictive power in a parsimonious way and allows unified
statistical inference about phase and amplitude components. The asymptotic
distribution of the estimators is derived and the finite-sample properties
are studied by simulation. An example of application involving ground-level
ozone trajectories is presented.

\end{abstract}

\section{Introduction}

The analysis of data consisting of curves or other types of functions,
rather than scalars or vectors, is increasingly common in statistics (Ramsay
\& Silverman, 2005). Many problems in this area involve modeling curves as
functions of other curves. For example, Figure \ref{fig:Sample_curves}(a)
shows daily trajectories of oxides of nitrogen in the city of Sacramento,
California, for 52 summer days in the year 2005, and Figure \ref%
{fig:Sample_curves}(b) shows the corresponding trajectories of ozone
concentration. The goal is to predict ozone concentration from oxides of
nitrogen.

\FRAME{ftbpFU}{6.2699in}{2.1352in}{0pt}{\Qcb{Ozone Example. Daily
trajectories of ground-level concentrations of (a) oxides of nitrogen and
(b) ozone in the city of Sacramento in the Summer of 2005.}}{\Qlb{%
fig:Sample_curves}}{sample_curves.eps}{\special{language "Scientific
Word";type "GRAPHIC";maintain-aspect-ratio TRUE;display "USEDEF";valid_file
"F";width 6.2699in;height 2.1352in;depth 0pt;original-width
10.9044in;original-height 3.4013in;cropleft "0.0767";croptop "1";cropright
"1";cropbottom "0";filename '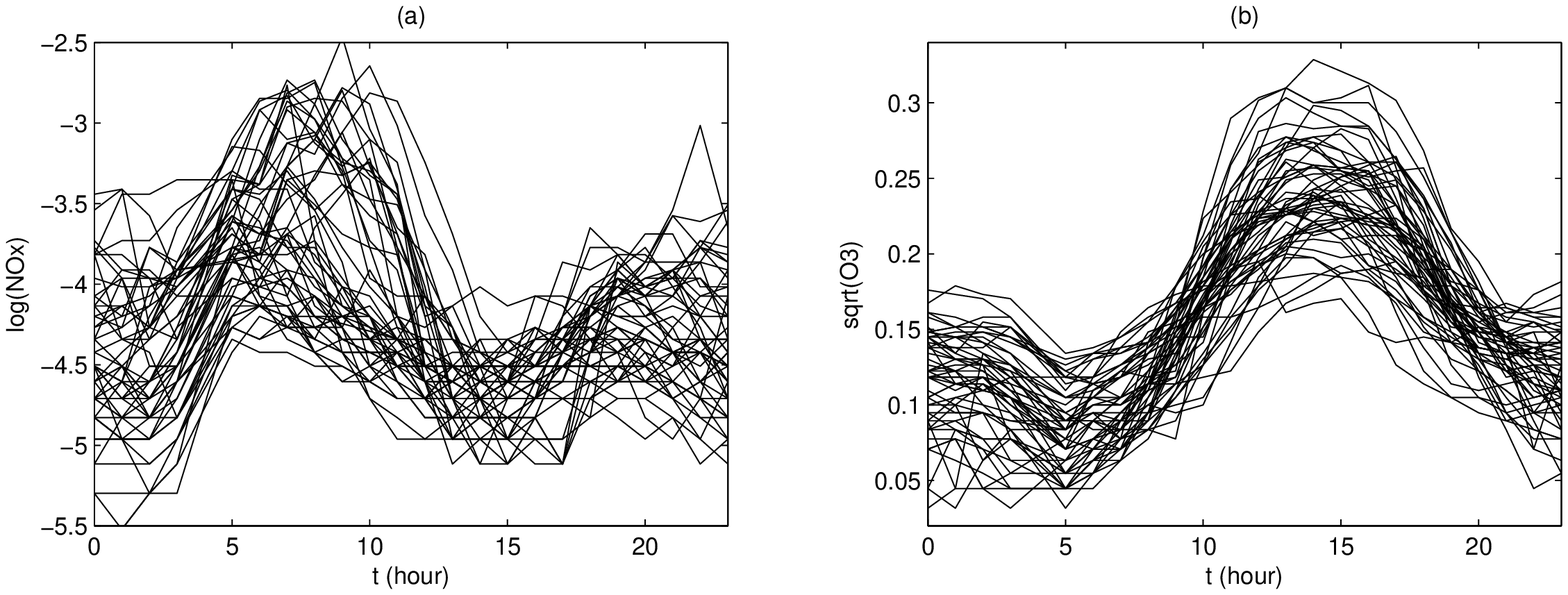';file-properties "XNPEU";}}

Functional linear regression models are normally used for this type of
problems (Ramsay \& Silverman, 2005, ch.~16). Recent papers have studied
different aspects of the functional linear regression model (Yao et al.,
2005; Cai \& Hall, 2006; Hall \& Horowitz, 2007; Crambes et al., 2009; James
et al., 2009). However, a characteristic feature of functional data that has
not been widely investigated in a regression context is phase variability.
Functional samples often present a few distinct features, such as peaks and
valleys, which vary in amplitude and location from curve to curve, as it is
clear in Figure \ref{fig:Sample_curves}. Functional linear regression is
usually based on functional principal components, which are well suited for
fitting amplitude variability but not for location or phase variability. It
may take an inordinate number of principal components to account even for
very basic phase-variability processes (Ramsay \& Silverman, 2005, ch.~7). A
more efficient strategy is to model amplitude and phase variability
separately: the former using traditional functional principal components and
the latter using warping models. This approach is more efficient, because
the combined model often provides a better fit with fewer parameters than
the classical principal component decomposition. It is also more
informative, because it provides direct information about the warping
process, which classical principal components only do indirectly. Several
warping methods have been proposed over the years (Gervini \& Gasser, 2004,
2005; James, 2007; Kneip et al., 2000; Kneip \& Ramsay, 2008; Liu \& M\"{u}%
ller, 2004; Ramsay \& Li, 1998; Tang \& M\"{u}ller, 2008, 2009; Wang \&
Gasser, 1999).

Common functional linear regression models inherit the problems of
functional principal components in presence of phase variability. Although a
high-dimensional model based on a large number of principal components can
provide a good fit to the data, the problem again is one of efficiency and
interpretability, not just minimizing prediction error. It is usually hard
to extract specific information about phase variability from a traditional
functional regression model because the two sources of variability, phase
and amplitude, are confounded in the model.

The curves shown in Figure \ref{fig:Sample_curves}, for example, show peaks
that vary not only in amplitude but also in location. It is reasonable to
hypothesize that a large peak in oxides of nitrogen will be followed by a
large peak in ozone concentration, and also that an early peak in oxides of
nitrogen will be followed by an early peak in ozone level, and vice-versa.
Perhaps there may also be an interaction between timing and amplitude of the
peaks. A common functional linear regression model of sufficiently high
dimension will be able to fit these data well from the point of view of
prediction error, but will not provide clear answers to these questions. A
regression model that explicitly incorporates a warping component and does
not confound the two sources of variability will be more useful for this,
and that is what we propose in this paper.

\section{\label{sec:Model}The Warped Functional Regression Model}

\subsection{Model specification}

Consider a sample of functions $(x_{1},y_{1})$, \ldots , $(x_{n},y_{n})$,
where $x_{i}(s)$ is the covariate and $y_{i}(t)$ the response, with $x_{i}:%
\mathcal{S}\rightarrow \mathbb{R}$ and $y_{i}:\mathcal{T}\rightarrow \mathbb{%
R}$, and $\mathcal{S}$ and $\mathcal{T}$ closed intervals in $\mathbb{R}$.
The functions $x_{i}(s)$ and $y_{i}(t)$ are usually not directly observable;
instead we observe discretizations of them, with added random noise, at time
grids $\{s_{ij}:j=1,\ldots ,\nu _{1i}\}$ and $\{t_{ij}:j=1,\ldots ,\nu
_{2i}\}$. Thus the observed data consist of vectors $(\mathbf{x}_{1},\mathbf{%
y}_{1}),\ldots ,(\mathbf{x}_{n},\mathbf{y}_{n})$, with $\mathbf{x}_{i}\in 
\mathbb{R}^{\nu _{1i}}$ and $\mathbf{y}_{i}\in \mathbb{R}^{\nu _{2i}}$ with
elements 
\begin{eqnarray}
x_{ij} &=&x_{i}(s_{ij})+\varepsilon _{ij},\ \ j=1,\ldots ,\nu _{1i},\
i=1,\ldots ,n,  \label{eq:raw_data_x} \\
y_{ij} &=&y_{i}(t_{ij})+\eta _{ij},\ \ j=1,\ldots ,\nu _{2i},\ i=1,\ldots ,n.
\label{eq:raw_data_y}
\end{eqnarray}%
We will assume that the measurement errors $\{\varepsilon _{ij}\}$ and $%
\{\eta _{ij}\}$ are independent with $\varepsilon _{ij}\sim N(0,\sigma
_{\varepsilon }^{2})$ and $\eta _{ij}\sim N(0,\sigma _{\eta }^{2})$.

The kind of curves we have in mind for our model will present a relatively
small number of peaks and valleys that systematically appear in all curves
but vary in amplitude and location. Then $\{x_{i}(s)\}$ and $\{y_{i}(t)\}$
can be thought of as compound processes 
\begin{eqnarray}
x_{i}(s) &=&x_{i}^{\ast }\{\omega _{i}^{-1}(s)\},  \label{eq:decomp_x} \\
y_{i}(t) &=&y_{i}^{\ast }\{\zeta _{i}^{-1}(t)\},  \label{eq:decomp_y}
\end{eqnarray}%
where $\{x_{i}^{\ast }(s)\}$ and $\{y_{i}^{\ast }(t)\}$ account for
amplitude variability and $\{\omega _{i}(s)\}$ and $\{\zeta _{i}(t)\}$
account for phase variability. The $\omega _{i}$s and the $\zeta _{i}$s are
monotone increasing warping functions with $\omega _{i}:\mathcal{S}%
\rightarrow \mathcal{S}$ and $\zeta _{i}:\mathcal{T}\rightarrow \mathcal{T}$%
. The aligned processes $\{x_{i}^{\ast }(s)\}$ and $\{y_{i}^{\ast }(t)\}$
follow principal-component decompositions 
\begin{eqnarray}
x_{i}^{\ast }(s) &=&\mu _{x}(s)+\sum_{k=1}^{p_{1}}u_{ik}\phi _{k}(s),
\label{eq:KL-model-x} \\
y_{i}^{\ast }(t) &=&\mu _{y}(t)+\sum_{l=1}^{p_{2}}v_{il}\psi _{l}(t),
\label{eq:KL-model-y}
\end{eqnarray}%
with $\left\{ \phi _{k}(s)\right\} $ and $\left\{ \psi _{l}(t)\right\} $
orthonormal functions in $L^{2}(\mathcal{S})$ and $L^{2}(\mathcal{T})$,
respectively, and $\{u_{ik}\}$ and $\{v_{il}\}$ uncorrelated zero-mean
random variables.

A few comments about (\ref{eq:decomp_x})--(\ref{eq:KL-model-y}) are in
order, because models (\ref{eq:decomp_x}) and (\ref{eq:decomp_y}) may seem
unidentifiable and models (\ref{eq:KL-model-x}) and (\ref{eq:KL-model-y})
may seem too restrictive for finite $p_{1}$ and $p_{2}$. These issues are
extensively discussed in Kneip \& Ramsay (2008, sec.~2.3) and in the
Supplementary Material. Proposition 1 in Kneip \& Ramsay (2008) shows that
if the $x_{i}$s have at most $K$ peaks and valleys and their derivatives $%
x_{i}^{\prime }(t)$ have at most $K$ zeros, then $x_{i}(t)$ admits the
decomposition $x_{i}(t)=\sum_{j=1}^{p}C_{ij}\xi _{j}\{v_{i}(t)\}$ for some $%
p\leq K+2$, where the $\xi _{j}$s are non-random basis functions, the $%
C_{ij} $s are random coefficients, and the $v_{i}$s are warping functions.
Orthogonalizing the $\xi _{j}$s one obtains model (\ref{eq:KL-model-x}).
Then $p_{1}$ in (\ref{eq:KL-model-x}) and $p_{2}$ in (\ref{eq:KL-model-y})
need not be large if the number of features to be aligned is small. The
identifiability of (\ref{eq:decomp_x}) and (\ref{eq:decomp_y}) given
amplitude models (\ref{eq:KL-model-x}) and (\ref{eq:KL-model-y}) and given
certain conditions on the warping family $\mathcal{W}$ is shown in the
Supplementary Material. If the summations in (\ref{eq:KL-model-x}) and (\ref%
{eq:KL-model-y}) were allowed to be infinite, then (\ref{eq:decomp_x}) and (%
\ref{eq:decomp_y}) would be unidentifiable. The practical effect of large $%
p_{1}$ and $p_{2}$ in (\ref{eq:KL-model-x}) and (\ref{eq:KL-model-y}) is
that the sample curves tend to present a large and unequal number of
features, and then it does not make sense to try to align them; in such
cases amplitude and phase variability essentially become indistinguishable.
Samples like that do occur in practice, but the methods we propose in this
paper are not intended for those situations.

The warping functions $\{\omega _{i}(s)\}$ and $\{\zeta _{i}(t)\}$ will be
modelled as monotone Hermite splines (Fritsch \& Carlson, 1980). Although
other families are possible, such as integrated splines (Ramsay, 1988),
monotone splines (Ramsay \& Li, 1998) and constrained B-splines (Brumback \&
Lindstrom, 2004), monotone Hermite splines are better suited for the
regression approach proposed here. Details about this family of warping
functions are given in Appendix \ref{app:Hermite}. We only mention here
that, like other spline families, this is a finite-dimensional
semiparametric family determined by a knot sequence chosen by the user.
Thus, the family $\{\omega _{i}(s)\}$ will be determined by a knot sequence $%
\mathbf{\tau }_{x0}=(\tau _{x01},\ldots ,\tau _{x0r_{1}})$ of strictly
increasing points in $\mathcal{S}$, and each $\omega _{i}(s)$ will be
determined by a corresponding sequence $\mathbf{\tau }_{xi}$ of basis
coefficients which satisfy $\omega _{i}(\tau _{x0j})=\tau _{xij}$ for $%
j=1,\ldots ,r_{1}$. Similarly, the family $\{\zeta _{i}(t)\}$ will be
determined by a knot sequence $\mathbf{\tau }_{y0}=(\tau _{y01},\ldots ,\tau
_{y0r_{2}})$ of strictly increasing points in $\mathcal{T}$ and each $\zeta
_{i}(t)$ will be determined by basis coefficients $\mathbf{\tau }_{yi}$
which satisfy $\zeta _{i}(\tau _{y0j})=\tau _{yij}$ for $j=1,\ldots ,r_{2}$.
The dual role of the $\mathbf{\tau }_{xi}$s and the $\mathbf{\tau }_{yi}$s
as basis coefficients and as values of $\omega _{i}(s)$ and $\zeta _{i}(t)$
at the knots is what makes Hermite splines appealing. It is natural then to
choose the knot sequences $\mathbf{\tau }_{x0}$ and $\mathbf{\tau }_{y0}$ to
roughly correspond to the average location of the main features of the $%
x_{i} $s and the $y_{i}$s. Like $p_{1}$ and $p_{2}$ in (\ref{eq:KL-model-x})
and (\ref{eq:KL-model-y}), the dimensions $r_{1}$ and $r_{2}$ need not be
large, since they will roughly correspond to the number of peaks and valleys
of the $x_{i}$s and the $y_{i}$s, which will not be large for the type of
applications we envision.

Unlike landmark registration, where the $\mathbf{\tau }_{xi}$s and the $%
\mathbf{\tau }_{yi}$s are individually estimated curve by curve, we will
treat the $\mathbf{\tau }_{xi}$s and the $\mathbf{\tau }_{yi}$s as latent
random effects, so they will not be estimated directly. This is a big
advantage in practice, since individual estimation of the $\mathbf{\tau }%
_{xi}$s and the $\mathbf{\tau }_{yi}$s is difficult when the number of
curves is large or when the curves are sparsely sampled. A minor
complication is that the $\mathbf{\tau }_{xi}$s and the $\mathbf{\tau }_{yi}$%
s are constrained to be monotone increasing in $\mathcal{S}$ and $\mathcal{T}
$, respectively, so for convenience we will work with their Jupp transforms $%
\mathbf{\theta }_{xi}$ and $\mathbf{\theta }_{yi}$ instead, which are
unconstrained vectors; the Jupp transform is defined in Appendix \ref%
{app:Hermite}.

Since the warping functions $\{\omega _{i}\}$ and $\{\zeta _{i}\}$ are
determined by the random effects $\mathbf{\theta }_{xi}$ and $\mathbf{\theta 
}_{yi}$, and the amplitude functions $\{x_{i}^{\ast }\}$ and $\{y_{i}^{\ast
}\}$ are determined by the random effects $\mathbf{u}_{i}$ and $\mathbf{v}%
_{i}$, we can specify an indirect regression model of the $y_{i}$s on the $%
x_{i}$s via the random effects: 
\begin{equation}
\left[ 
\begin{array}{c}
\mathbf{v}_{i} \\ 
\mathbf{\theta }_{yi}%
\end{array}%
\right] =\left[ 
\begin{array}{c}
\mathbf{0} \\ 
\mathbf{\theta }_{y0}%
\end{array}%
\right] +\mathbf{A}\left( \left[ 
\begin{array}{c}
\mathbf{u}_{i} \\ 
\mathbf{\theta }_{xi}%
\end{array}%
\right] -\left[ 
\begin{array}{c}
\mathbf{0} \\ 
\mathbf{\theta }_{x0}%
\end{array}%
\right] \right) +\mathbf{e}_{i},  \label{eq:Reg-model}
\end{equation}%
where $\mathbf{A}$ is the $\left( p_{2}+r_{2}\right) \times \left(
p_{1}+r_{1}\right) $ regression matrix and $\mathbf{e}_{i}$ is an error
term, which we assume $N(\mathbf{0},\mathbf{\Sigma }_{e})$ with $\mathbf{%
\Sigma }_{e}$ diagonal. For interpretability we split $\mathbf{A}$ into four
blocks corresponding to $\mathbf{u}_{i}$, $\mathbf{\theta }_{xi}$, $\mathbf{v%
}_{i}$ and $\mathbf{\theta }_{yi}$: 
\[
\mathbf{A}=\left[ 
\begin{array}{cc}
\mathbf{A}_{11} & \mathbf{A}_{12} \\ 
\mathbf{A}_{21} & \mathbf{A}_{22}%
\end{array}%
\right] , 
\]%
with $\mathbf{A}_{11}\in \mathbb{R}^{p_{2}\times p_{1}}$, $\mathbf{A}%
_{12}\in \mathbb{R}^{p_{2}\times r_{1}}$, $\mathbf{A}_{21}\in \mathbb{R}%
^{r_{2}\times p_{1}}$ and $\mathbf{A}_{22}\in \mathbb{R}^{r_{2}\times r_{1}}$%
. Then (\ref{eq:KL-model-x}), (\ref{eq:KL-model-y}) and (\ref{eq:Reg-model})
imply 
\begin{eqnarray}
y_{i}^{\ast }(t)-\mu _{y}(t) &=&\int \beta (s,t)\{x_{i}^{\ast }(s)-\mu
_{x}(s)\}\ \mathrm{ds}+\mathbf{\gamma }_{1}(t)^{T}(\mathbf{\theta }_{xi}-%
\mathbf{\theta }_{x0})+\delta _{i}(t),  \label{eq:Reg-model-y} \\
\mathbf{\theta }_{yi}-\mathbf{\theta }_{y0} &=&\int \mathbf{\gamma }%
_{2}(s)\{x_{i}^{\ast }(s)-\mu _{x}(s)\}\ \mathrm{ds}+\mathbf{A}_{22}(\mathbf{%
\theta }_{xi}-\mathbf{\theta }_{x0})+\mathbf{e}_{i2},
\label{eq:Reg-model-theta}
\end{eqnarray}%
where $\beta (s,t)=\mathbf{\psi }(t)^{T}\mathbf{A}_{11}\mathbf{\phi }(s)$, $%
\mathbf{\gamma }_{1}(t)^{T}=\mathbf{\psi }(t)^{T}\mathbf{A}_{12}$, $\mathbf{%
\gamma }_{2}(s)=\mathbf{A}_{21}\mathbf{\phi }(s)$ and $\delta _{i}(t)=%
\mathbf{\psi }(t)^{T}\mathbf{e}_{i1}$. Thus, for example, $\mathbf{A}_{12}=%
\mathbf{0}$ implies that $\mathbf{\gamma }_{1}(t)=\mathbf{0}$ and then the
amplitude variability of the responses is unrelated to the time variability
of the covariates; similarly, $\mathbf{A}_{21}=\mathbf{0}$ implies that $%
\mathbf{\gamma }_{2}(s)=\mathbf{0}$ and then the time variability of the
responses is unrelated to the amplitude variability of the covariates.

\subsection{\label{sec:Estimation}Estimation and prediction}

Models (\ref{eq:KL-model-x}) and (\ref{eq:KL-model-y}) depend on functional
parameters that need to be estimated: the mean functions $\mu _{x}(s)$ and $%
\mu _{y}(t)$ and the principal components $\{\phi _{k}(s)\}$ and $\{\psi
_{l}(t)\}$. We will do that via B-splines. Let $\mathbf{b}%
_{x}(s)=(b_{x1}(s),\ldots ,b_{xq_{1}}(s))^{T}$ be a B-spline basis in $L^{2}(%
\mathcal{S})$ and $\mathbf{b}_{y}(t)=(b_{y1}(t),\ldots ,b_{yq_{2}}(t))^{T}$
a B-spline basis in $L^{2}(\mathcal{T})$. Let $\mu _{x}(s)=\mathbf{b}%
_{x}^{T}(s)\mathbf{m}_{x}$, $\mu _{y}(t)=\mathbf{b}_{y}^{T}(t)\mathbf{m}_{y}$%
, $\phi _{k}(s)=\mathbf{b}_{x}^{T}(s)\mathbf{c}_{k}$ and $\psi _{l}(t)=%
\mathbf{b}_{y}^{T}(t)\mathbf{d}_{l}$, for $\mathbf{m}_{x}\in \mathbb{R}%
^{q_{1}}$, $\mathbf{m}_{y}\in \mathbb{R}^{q_{2}}$, $\mathbf{c}_{k}\in 
\mathbb{R}^{q_{1}}$ and $\mathbf{d}_{l}\in \mathbb{R}^{q_{2}}$. The
orthogonality restrictions on the $\phi _{k}$s and the $\psi _{l}$s can be
expressed as $\mathbf{C}^{T}\mathbf{J}_{x}\mathbf{C}=\mathbf{I}_{p_{1}}$ and 
$\mathbf{D}^{T}\mathbf{J}_{y}\mathbf{D}=\mathbf{I}_{p_{2}}$, where $\mathbf{%
C=[c}_{1},\ldots ,\mathbf{c}_{p_{1}}]\in \mathbb{R}^{q_{1}\times p_{1}}$, $%
\mathbf{D}=[\mathbf{d}_{1},\ldots ,\mathbf{d}_{p_{2}}]\in \mathbb{R}%
^{q_{2}\times p_{2}}$, $\mathbf{J}_{x}=\int \mathbf{b}_{x}(s)\mathbf{b}%
_{x}^{T}(s)\mathrm{d}s$ and $\mathbf{J}_{y}=\int \mathbf{b}_{y}(t)\mathbf{b}%
_{y}^{T}(t)\mathrm{d}t$.

If the curves $\{x_{i}\}$ and $\{y_{i}\}$ were observed on dense time grids
and individual smoothing were possible, the spline coefficients and the rest
of the model parameters could be estimated by least squares. However, we are
more interested in applications where the trajectories are not densely
sampled. Then we will treat $\mathbf{u}_{i}$, $\mathbf{v}_{i}$, $\mathbf{%
\theta }_{xi}$ and $\mathbf{\theta }_{yi}$ as latent variables and estimate
the model parameters by maximum likelihood. We assume $\mathbf{w}_{i}=(%
\mathbf{u}_{i}^{T},\mathbf{\theta }_{xi}^{T})^{T}$ is jointly multivariate
Normal of dimension $d_{1}=p_{1}+r_{1}$, with mean and covariance given by 
\[
\mathbf{\mu }_{w}=\left[ 
\begin{array}{c}
\mathbf{0} \\ 
\mathbf{\theta }_{x0}%
\end{array}%
\right] ,\ \ \mathbf{\Sigma }_{w}=\left[ 
\begin{array}{cc}
\mathbf{\Lambda } & \mathbf{\Sigma }_{u\theta _{x}} \\ 
\mathbf{\Sigma }_{u\theta _{x}}^{T} & \mathbf{\Sigma }_{\theta _{x}}%
\end{array}%
\right] , 
\]%
where $\mathbf{\theta }_{x0}$ the Jupp transform of the knot vector $\mathbf{%
\tau }_{x0}$ and $\mathbf{\Lambda }=\mathrm{diag}(\lambda _{1},\ldots
,\lambda _{p_{1}})$. This and model (\ref{eq:Reg-model}) imply that $\mathbf{%
z}_{i}=(\mathbf{v}_{i}^{T},\mathbf{\theta }_{yi}^{T})^{T}$ is multivariate
Normal of dimension $d_{2}=p_{2}+r_{2}$ with mean and covariance given by 
\[
\mathbf{\mu }_{z}=\left[ 
\begin{array}{c}
\mathbf{0} \\ 
\mathbf{\theta }_{y0}%
\end{array}%
\right] ,\ \ \mathbf{\Sigma }_{z}=\mathbf{A\Sigma }_{w}\mathbf{A}^{T}+%
\mathbf{\Sigma }_{e}, 
\]%
where $\mathbf{\theta }_{y0}$ is the Jupp transform of the knot vector $%
\mathbf{\tau }_{y0}$. Thus $\mathbf{v}_{i}\sim N(\mathbf{0},\mathbf{\Gamma }%
) $ with $\mathbf{\Gamma }=\mathbf{A}_{1\cdot }\mathbf{\Sigma }_{w}\mathbf{A}%
_{1\cdot }^{T}+\mathbf{\Sigma }_{e,11}$, where $\mathbf{A}_{1\cdot }=[%
\mathbf{A}_{11},\mathbf{A}_{12}]$ and $\mathbf{\Sigma }_{e,11}$ the $%
p_{2}\times p_{2}$ upper-left diagonal block of $\mathbf{\Sigma }_{e}$.
Since $\mathbf{\Gamma }$ has to be diagonal by model (\ref{eq:KL-model-y}),
and $\mathbf{\Sigma }_{e}$ was assumed diagonal, it follows that $\mathbf{A}%
_{1\cdot }\mathbf{\Sigma }_{w}\mathbf{A}_{1\cdot }^{T}$ must be diagonal,
which imposes an additional restriction on the parameters.

To summarize, the parameters of this model are: the regression matrix $%
\mathbf{A}$, the residual covariance matrix $\mathbf{\Sigma }_{e}$, the
covariance matrix $\mathbf{\Sigma }_{w}$ of the explanatory random effects $%
\mathbf{w}_{i}$, the spline coefficients $\mathbf{m}_{x}$, $\mathbf{m}_{y}$, 
$\mathbf{C}$ and $\mathbf{D}$ of the functional parameters, and the
variances $\sigma _{\varepsilon }^{2}$ and $\sigma _{\eta }^{2}$ of the
random noise in (\ref{eq:raw_data_x}) and (\ref{eq:raw_data_y}). The
derivation of the likelihood function and the EM algorithm to compute these
estimators are discussed in Appendix \ref{app:Likelihood} and in the
Supplementary Material.

In addition to the model parameters there are meta-parameters that need to
be chosen by the user, such as the dimension and knot placement of the
B-spline bases for the functional parameters. This can be done either
subjectively or by cross-validation. Since the method `borrows strength'
across curves, it is possible to use a larger number of knots than would be
practical for single-curve smoothing. The other meta-parameters that need to
be specified are the number of components in models (\ref{eq:KL-model-x})
and (\ref{eq:KL-model-y}), $p_{1}$ and $p_{2}$, and the warping dimensions $%
r_{1}$ and $r_{2}$. As already discussed, these quantities should roughly
correspond to the number of salient features of the $x_{i}$s and the $y_{i}$%
s.

In addition to parameter estimation, it is usually of interest to predict a
response curve for a given covariate curve. This can be done in a
straightforward way. Given a covariate data vector $\mathbf{x}_{n+1}$,
obtained by discretizing a covariate curve $x_{n+1}(s)$ on some time grid,
the predictors $\mathbf{\hat{v}}_{n+1}$ and $\mathbf{\hat{\theta}}_{y,n+1}$
of the response random effects are given by $\hat{E}(\mathbf{v}_{n+1}|%
\mathbf{x}_{n+1})$ and $\hat{E}(\mathbf{\theta }_{y,n+1}|\mathbf{x}_{n+1})$,
which under model (\ref{eq:Reg-model}) come down to $\mathbf{\hat{v}}_{n+1}=%
\mathbf{\hat{A}}_{11}\hat{E}(\mathbf{u}_{n+1}|\mathbf{x}_{n+1})+\mathbf{\hat{%
A}}_{12}\{\hat{E}(\mathbf{\theta }_{x,n+1}|\mathbf{x}_{n+1})-\mathbf{\theta }%
_{x0}\}$ and $\mathbf{\hat{\theta}}_{y,n+1}=\mathbf{\hat{A}}_{21}\hat{E}(%
\mathbf{u}_{n+1}|\mathbf{x}_{n+1})+\mathbf{\hat{A}}_{22}\{\hat{E}(\mathbf{%
\theta }_{x,n+1}|\mathbf{x}_{n+1})-\mathbf{\theta }_{x0}\}$. With $\mathbf{%
\hat{v}}_{n+1}$ and $\mathbf{\hat{\theta}}_{y,n+1}$ we compute $\hat{y}%
_{n+1}^{\ast }(t)$ and $\hat{\zeta}_{n+1}(t)$ respectively, and then $\hat{y}%
_{n+1}(t)=\hat{y}_{n+1}^{\ast }\{\hat{\zeta}_{n+1}^{-1}(t)\}$.

\section{\label{sec:Inference}Inference}

Consider now the asymptotic distribution of $\mathbf{\hat{A}}$ when the
number of curves $n$ goes to infinity. For simplicity, we will assume that
the time grids are equal for all individuals, which makes the raw data
vectors $(\mathbf{x}_{1},\mathbf{y}_{1}),\ldots ,(\mathbf{x}_{n},\mathbf{y}%
_{n})$ independent and identically distributed. We will also assume that the
functional parameters belong to the spline space used for estimation, whose
dimension is held fixed.

The asymptotic analysis is not entirely straightforward due to the parameter
constraints. For this reason we will use the results of Geyer (1994). Since
we are only interested in the marginal asymptotic distribution of $\mathbf{%
\hat{A}}$ and not in the asymptotic covariance between $\mathbf{\hat{A}}$
and the rest of the parameters, we can assume without loss of generality
that $\mathbf{\Sigma }_{e}$, $\mathbf{m}_{x}$, $\mathbf{m}_{y}$, $\mathbf{C}$%
, $\mathbf{D}$, $\sigma _{\varepsilon }^{2}$ and $\sigma _{\eta }^{2}$ are
fixed and known, because this assumption does not alter the asymptotic
covariance matrix of $\mathbf{\hat{A}}$. However, in principle we cannot
assume that $\mathbf{\Sigma }_{w}$ is fixed and known because $\mathbf{%
\Sigma }_{w}$ is part of the condition that $\mathbf{A}_{1\cdot }\mathbf{%
\Sigma }_{w}\mathbf{A}_{1\cdot }^{T}$\ be diagonal. So we will derive the
joint asymptotic distribution of $\mathbf{\hat{A}}$ and $\mathbf{\hat{\Sigma}%
}_{w}$, even though we are only interested in the marginal distribution of $%
\mathbf{\hat{A}}$.

The parameter of interest is then, in vector form, 
\begin{equation}
\mathbf{\zeta }=\left[ 
\begin{array}{c}
\mathrm{vec}(\mathbf{A}^{T}) \\ 
\mathrm{v}(\mathbf{\Sigma }_{w})%
\end{array}%
\right] ,  \label{eq:theta}
\end{equation}%
where $\mathrm{v}(\mathbf{\Sigma }_{w})$ denotes the $\mathrm{vec}$ of the
lower-triangular part of $\mathbf{\Sigma }_{w}$, including the diagonal. The
dimension of $\mathbf{\zeta }$ is then $d=d_{1}d_{2}+d_{1}(d_{1}+1)/2$. The
restriction that $\mathbf{A}_{1\cdot }\mathbf{\Sigma }_{w}\mathbf{A}_{1\cdot
}^{T}$ be diagonal can be expressed as a system of $m=(p_{2}-1)p_{2}/2$
constraints of the form $h_{ij}(\mathbf{\zeta })=0$, where $h_{ij}(\mathbf{%
\zeta })=\mathbf{a}_{i}^{T}\mathbf{\Sigma }_{w}\mathbf{a}_{j}$ and $\mathbf{a%
}_{i}^{T}$ is the $i$th row of $\mathbf{A}$. The functions $h_{ij}$ can be
stacked together into a single vector-valued function $\mathbf{h}:\mathbb{R}%
^{d}\rightarrow \mathbb{R}^{m}$, and the constrained parameter space can be
expressed as $C=\left\{ \mathbf{\zeta }\in \mathbb{R}^{d}:\mathbf{h}(\mathbf{%
\zeta })=\mathbf{0}\right\} $. The additional condition that $\mathbf{\Sigma 
}_{w}$ be positive definite does not alter the asymptotic distribution of
the estimator because $\mathbf{\Sigma }_{w}$ lies in the interior of this
space, not on the border. Let $\mathbf{\zeta }_{0}$ be the true value of the
parameter $\mathbf{\zeta }$. Since $\mathbf{h}(\mathbf{\zeta })$ is
continuously differentiable, the tangent cone of $C$ at $\mathbf{\zeta }_{0}$
is $T_{C}(\mathbf{\zeta }_{0})=\left\{ \mathbf{\delta }\in \mathbb{R}^{d}:%
\mathrm{D}\mathbf{h}(\mathbf{\zeta }_{0})\mathbf{\delta }=\mathbf{0}\right\} 
$, where $\mathrm{D}$ is the differential (Rockafellar \& Wets, 1998,
ch.~6.B). The asymptotic distribution of the constrained estimator $\mathbf{%
\hat{\zeta}}_{n}$ is simple in this case: it is just the usual asymptotic
Normal distribution of an unconstrained maximum likelihood estimator,
projected on $T_{C}(\mathbf{\zeta }_{0})$.

Specifically, let 
\begin{eqnarray}
\mathbf{M}(\mathbf{x,y}) &=&E\{(\mathbf{w-\mu }_{w})(\mathbf{w-\mu }%
_{w})^{T}|(\mathbf{x,y})\},  \label{eq:M} \\
\mathbf{N}(\mathbf{x,y}) &=&E\{(\mathbf{w-\mu }_{w})(\mathbf{z-\mu }%
_{z})^{T}|(\mathbf{x,y})\},  \label{eq:N}
\end{eqnarray}%
and 
\begin{equation}
\mathbf{U(x,y)}=\left[ 
\begin{array}{c}
\mathrm{vec}\{\mathbf{N(x,y)\Sigma }_{e,0}^{-1}\}-\mathrm{vec}\{\mathbf{%
M(x,y)A}_{0}^{T}\mathbf{\Sigma }_{e,0}^{-1}\} \\ 
(-1/2)\mathbf{D}_{d_{1}}^{T}\mathrm{vec}\{\mathbf{\Sigma }_{w,0}^{-1}-%
\mathbf{\Sigma }_{w,0}^{-1}\mathbf{M(x,y)\Sigma }_{w,0}^{-1}\}%
\end{array}%
\right] ,  \label{eq:U}
\end{equation}%
where $\mathbf{D}_{d_{1}}$ is the duplication matrix that satisfies $\mathrm{%
vec}(\mathbf{\Sigma }_{w})=\mathbf{D}_{d_{1}}\mathrm{v}(\mathbf{\Sigma }%
_{w}) $ (Magnus \& Neudecker, 1999, ch.~3). It is shown in the Supplementary
Material that $\mathbf{U(x,y)}$ is the likelihood score function $\nabla _{%
\mathbf{\zeta }}\log f(\mathbf{x,y;\zeta })$ at $\mathbf{\zeta =\zeta }_{0}$%
. Let $\mathbf{B}=\mathrm{D}\mathbf{h}(\mathbf{\zeta }_{0})$, which is an $%
m\times d$ matrix of rank $m$ with rows 
\[
\nabla h_{ij}(\mathbf{\zeta })^{T}=[\mathbf{a}_{i}^{T}\mathbf{\Sigma }_{w}(%
\mathbf{e}_{j}\otimes \mathbf{I}_{d_{1}})+\mathbf{a}_{j}^{T}\mathbf{\Sigma }%
_{w}(\mathbf{e}_{i}\otimes \mathbf{I}_{d_{1}}),\mathbf{0}_{r_{2}d_{1}}^{T},(%
\mathbf{a}_{j}^{T}\otimes \mathbf{a}_{i}^{T})\mathbf{D}_{d_{1}}], 
\]%
where $\mathbf{e}_{i}$ is the $i$th canonical vector in $\mathbb{R}^{p_{2}}$%
. Let $\mathbf{\Xi }$ be an orthogonal $d\times (d-m)$ matrix of rank $d-m$
such that $\mathbf{B\Xi }=\mathbf{0}$, which can be computed for instance
via the singular value decomposition of the orthogonal projector $\mathbf{I}%
_{d}-\mathbf{B}^{T}(\mathbf{BB}^{T})^{-1}\mathbf{B}$; this matrix is not
unique but Theorem \ref{theo:Asymp} below is invariant under the choice of $%
\mathbf{\Xi }$.

\begin{theorem}
\label{theo:Asymp}Under the above conditions, the asymptotic distribution of 
$\sqrt{n}(\mathbf{\hat{\zeta}}_{n}-\mathbf{\zeta }_{0})$ is $N\{\mathbf{0},%
\mathbf{\Xi }(\mathbf{\Xi }^{T}\mathbf{V\Xi })^{-1}\mathbf{\Xi }^{T}\}$
where $\mathbf{V}=E\{\mathbf{U(x,y)U(x,y)}^{T}\}$.
\end{theorem}

Matrix $\mathbf{V}$ in Theorem \ref{theo:Asymp} is Fisher's Information
Matrix for this model and can be estimated by 
\begin{equation}
\mathbf{\hat{V}}_{n}=\frac{1}{n}\sum_{i=1}^{n}\mathbf{\hat{U}(x}_{i}\mathbf{%
,y}_{i}\mathbf{)\hat{U}(x}_{i}\mathbf{,y}_{i}\mathbf{)}^{T},  \label{eq:V}
\end{equation}%
where the `hat'\ in $\mathbf{U}$ denotes that the true parameters in (\ref%
{eq:U}) are replaced by their estimators. The proof of Theorem \ref%
{theo:Asymp} is given in the Appendix.

The assumption that the time grids were equal for all individuals was a
simplification to make the data vectors $(\mathbf{x}_{i},\mathbf{y}_{i})$,
and consequently the likelihood scores (\ref{eq:U}), identically
distributed. In many applications, however, this will not be the case and
the time grids will be unequal, giving $\mathbf{x}_{i}\in \mathbb{R}^{\nu
_{1i}}$ and $\mathbf{y}_{i}\in \mathbb{R}^{\nu _{2i}}$ which are still
independent but not identically distributed due to the different dimensions.
Usually this does not affect the final asymptotic result as long as (\ref%
{eq:V}) does not become degenerate, as shown for instance by Pollard (1990,
ch.~11) in the context of regression with non-random covariates. Although
the Fisher Information Matrix $\mathbf{V}$ as such does not exists, (\ref%
{eq:M}) and (\ref{eq:N}) and consequently (\ref{eq:U}) and (\ref{eq:V}) can
still be computed with $(\mathbf{x}_{i},\mathbf{y}_{i})$s of unequal
dimensions. The statement of Theorem \ref{theo:Asymp} should then be
re-expressed as 
\begin{equation}
\sqrt{n}\{\mathbf{\Xi }(\mathbf{\Xi }^{T}\mathbf{\hat{V}}_{n}\mathbf{\Xi }%
)^{-1}\mathbf{\Xi }^{T}\}^{-1/2}(\mathbf{\hat{\zeta}}_{n}-\mathbf{\zeta }%
_{0})\longrightarrow N(\mathbf{0},\mathbf{I}_{d})  \label{eq:asymp_uneq}
\end{equation}%
in distribution.

\section{\label{sec:Simulations}Simulations}

\subsection{\label{subsec:Sim_estim_acc}Estimation accuracy}

To study the finite-sample accuracy of the proposed estimators we simulated
data from the following models:

\begin{itemize}
\item Model 1: a one-dimensional amplitude and warping model, with $\mu
_{x}(s)=.6\varphi (s,.3,.1)+.4\varphi (s,.6,.1)$, $\phi _{1}(s)=\varphi
(s,.3,.1)/1.6796$, $\mu _{y}(t)=.6\varphi (t,.5,.1)+.4\varphi (t,.8,.1)$ and 
$\psi _{1}(t)=\varphi (t,.5,.1)/1.6796$, for $s$ and $t$ in $[0,1]$, where $%
\varphi (s,\mu ,\sigma )$ denotes the $N(\mu ,\sigma ^{2})$ density
function. The warping functions followed Hermite spline models with knots $%
\tau _{x0}=.3$ and $\tau _{y0}=.5$. Thus, although $\mu _{x}(s)$ and $\mu
_{y}(t)$ have two peaks, phase and amplitude variability are concentrated on
the main peak. The regression matrix $\mathbf{A}$ was the identity matrix,
so there was no relationship between covariate phase variability and
response amplitude variability, or vice versa, in this model. The other
parameters were $\mathbf{\Sigma }_{w}=\mathrm{diag}(.2^{2},.1^{2})$, $%
\mathbf{\Sigma }_{e}=.07^{2}\mathbf{I}_{2}$, and $\sigma _{\varepsilon
}=\sigma _{\eta }=.05$.

\item Model 2: same as Model 1 but with a non-diagonal $\mathbf{A}$;
specifically, $a_{11}=a_{22}=1$ and $a_{12}=a_{21}=.5$, so there was a
relationship between covariate phase variability and response amplitude
variability, and vice versa, in this model.

\item Model 3: a two-dimensional amplitude and warping model, with $\mu
_{x}(s)$, $\mu _{y}(t)$, $\phi _{1}(s)$ and $\psi _{1}(t)$ as in Model 1, $%
\phi _{2}(s)$ the function $\varphi (s,.6,.1)$ orthogonalized with $\phi
_{1}(s)$, and $\psi _{1}(t)$ the function $\varphi (t,.8,.1)$ orthogonalized
with $\psi _{1}(t)$. The warping functions followed Hermite spline models
with knots $\mathbf{\tau }_{x0}=(.3,.6)$ and $\mathbf{\tau }_{y0}=(.5,.8)$.
This model, then, has amplitude and phase variability at both peaks of $\mu
_{x}(s)$ and $\mu _{y}(t)$. The regression matrix $\mathbf{A}$ was the
identity, and the other parameters were $\mathbf{\Sigma }_{w}=\mathrm{diag}%
(.2^{2},.1^{2},.1^{2},.1^{2})$, $\mathbf{\Sigma }_{e}=.07^{2}\mathbf{I}_{4}$%
, and $\sigma _{\varepsilon }=\sigma _{\eta }=.05$.

\item Model 4: same as Model 3 but with a non-diagonal regression matrix $%
\mathbf{A}$, with blocks $\mathbf{A}_{11}=\mathbf{A}_{22}=\mathbf{I}_{2}$
and $\mathbf{A}_{12}=\mathbf{A}_{21}=.5\mathbf{I}_{2}$.

\item Model 5: a one-dimensional amplitude model like Model 1 but with
warping functions that do not follow a regression model and do not belong to
the Hermite-spline family; they belonged to a generic B-spline family with
monotone increasing coefficients, which produces monotone increasing
functions (Brumback \& Lindstrom, 2004). Specifically, if $\mathbf{b}(s)$
are cubic B-splines with 7 equally-spaced knots in $(0,1)$ and $\mathbf{c}%
_{0}$ is such that $\mathbf{b}(s)^{T}\mathbf{c}_{0}\equiv s$, the identity,
then we generated $\mathbf{c}_{i}\sim N(\mathbf{c}_{0},.05^{2}\mathbf{I}%
_{9}) $ and took $\omega
_{i}^{-1}(s)=\{g_{i}(s)-g_{i}(0)\}/\{g_{i}(1)-g_{i}(0)\}$, with $g_{i}(s)=%
\mathbf{b}(s)^{T}\mathbf{c}_{(i)}$ and $\mathbf{c}_{(i)}$ the coefficients
of $\mathbf{c}_{i}$ sorted in increasing order. The inverse warping
functions of the responses, the $\zeta _{i}^{-1}(t)$s, were generated in an
analogous way and were independent of the $\omega _{i}^{-1}(s)$s.

\item Model 6: a two-dimensional amplitude model like Model 3 with a
non-Hermite warping model like Model 5.
\end{itemize}

Two sample sizes, $n=50$ and $n=100$, were considered for each model. Each
scenario was replicated 500 times. In all cases the time grids $%
\{s_{i1},\ldots ,s_{i\nu _{1i}}\}$ and $\{t_{i1},\ldots ,t_{i\nu _{2i}}\}$
were random and irregular, with $\nu _{1i}$ and $\nu _{2i}$ uniformly
distributed between 10 and 20, and independent of one another, and $s_{ij}$
and $t_{ij}$ uniformly distributed on $[0,1]$.

For each sample we computed the proposed warped functional regression
estimator using cubic B-splines with 10 equally spaced knots for the
functional parameters, with the number of principal components $p_{1}$ and $%
p_{2}$ equal to the true model quantities, that is, $p_{1}=p_{2}=1$ for
Models 1, 2 and 5, and $p_{1}=p_{2}=2$ for Models 3, 4 and 6. The
specification of the warping functions, although always in a Hermite-spline
family, varied from model to model. For Models 1 and 2 we used the same
family used for estimation. For Models 3 and 4, however, we used
Hermite-spline families with single knots at $\tau _{x0}=.45$ and $\tau
_{y0}=.65$, so as to study the behavior of the estimator when the number of
warping knots is underspecified. For Model 5 we used Hermite splines with
knots at $\mathbf{\tau }_{x0}=(.3,.6)$ and $\mathbf{\tau }_{y0}=(.5,.8)$,
and for Model 6 we used Hermite splines with knots at $\tau _{x0}=.45$ and $%
\tau _{y0}=.65$; this allows us to study the advantages of doing some kind
of warping as opposed to not doing any warping at all, since the true
warping processes of Models 5 and 6 do not follow a regression model and do
not belong to the Hermite spline family.

For comparison we also computed ordinary functional regression estimators
based on principal components, as in e.g.~M\"{u}ller et al.~(2008), with the
difference that the principal components were computed by maximum likelihood
via B-spline models, as in James et al.~(2000), rather than by kernel
smoothing.

As measures of performance we computed bias and root mean squared errors of $%
\hat{\beta}(s,t)$, $\hat{\mu}_{x}(s)$, $\hat{\mu}_{y}(t)$, $\{\hat{\phi}%
_{j}(s)\}$ and $\{\hat{\psi}_{j}(t)\}$. We defined as `bias' of $\hat{\mu}%
_{x}$ the quantity $(\int [E\{\hat{\mu}_{x}(s)\}-\mu _{x}(s)]^{2}\mathrm{ds}%
)^{1/2}$ and as `root mean squared error' the quantity $(\int E[\{\hat{\mu}%
_{x}(s)-\mu _{x}(s)\}^{2}]\mathrm{ds})^{1/2}$. For $\hat{\mu}_{y}(t)$ and $%
\hat{\beta}(s,t)$ the definitions were analogous, with double integrals for
the latter. For the principal component estimators, which have undefined
signs, we actually computed the bias and root mean squared errors of the
bivariate functions $\hat{\phi}_{j}(s)\hat{\phi}_{j}(s^{\prime })$ and $\hat{%
\psi}_{j}(t)\hat{\psi}_{j}(t^{\prime })$, which are sign-invariant. These
are reported in Tables \ref{tab:Simulations_1_50} and \ref%
{tab:Simulations_1_100}; for $\hat{\mu}_{x}$ and $\hat{\mu}_{y}$ the
quantities have been multiplied by 10 to eliminate leading zeros.

\begin{table}[tbp] \centering%

\noindent 
\begin{tabular}{cccccccccc}
& \multicolumn{4}{c}{Model 1} &  & \multicolumn{4}{c}{Model 2} \\ 
& \multicolumn{2}{c}{bias} & \multicolumn{2}{c}{rmse} &  & 
\multicolumn{2}{c}{bias} & \multicolumn{2}{c}{rmse} \\ 
Param. & W & O & W & O &  & W & O & W & O \\ 
$\beta $ & 0.12 & 0.19 & 0.21 & 0.30 &  & 0.11 & 0.69 & 0.33 & 0.74 \\ 
$\mu _{x}$ & 0.10 & 0.19 & 0.34 & 0.37 &  & 0.12 & 0.19 & 0.38 & 0.37 \\ 
$\mu _{y}$ & 0.13 & 0.32 & 0.42 & 0.51 &  & 0.16 & 0.59 & 0.49 & 0.73 \\ 
$\phi _{1}$ & 0.05 & 0.06 & 0.15 & 0.18 &  & 0.08 & 0.05 & 0.23 & 0.18 \\ 
$\psi _{1}$ & 0.15 & 0.21 & 0.22 & 0.34 &  & 0.09 & 0.83 & 0.20 & 0.85 \\ 
&  &  &  &  &  &  &  &  &  \\ 
& \multicolumn{4}{c}{Model 3} &  & \multicolumn{4}{c}{Model 4} \\ 
$\beta $ & 0.37 & 1.00 & 1.15 & 1.14 &  & 0.47 & 1.23 & 1.39 & 1.32 \\ 
$\mu _{x}$ & 0.14 & 0.27 & 0.46 & 0.47 &  & 0.13 & 0.26 & 0.47 & 0.46 \\ 
$\mu _{y}$ & 0.16 & 0.38 & 0.56 & 0.58 &  & 0.19 & 0.65 & 0.61 & 0.81 \\ 
$\phi _{1}$ & 0.92 & 0.99 & 1.23 & 1.40 &  & 0.96 & 0.99 & 1.36 & 1.40 \\ 
$\phi _{2}$ & 0.25 & 0.93 & 0.59 & 1.06 &  & 0.22 & 0.96 & 0.58 & 1.07 \\ 
$\psi _{1}$ & 0.99 & 0.99 & 1.40 & 1.40 &  & 0.99 & 0.99 & 1.40 & 1.39 \\ 
$\psi _{2}$ & 0.17 & 0.87 & 0.47 & 1.21 &  & 0.20 & 0.62 & 0.48 & 1.03 \\ 
&  &  &  &  &  &  &  &  &  \\ 
& \multicolumn{4}{c}{Model 5} &  & \multicolumn{4}{c}{Model 6} \\ 
$\beta $ & 0.18 & 0.73 & 0.73 & 0.78 &  & 0.80 & 1.05 & 1.56 & 1.11 \\ 
$\mu _{x}$ & 0.44 & 0.94 & 0.84 & 1.10 &  & 0.55 & 0.94 & 0.93 & 1.11 \\ 
$\mu _{y}$ & 0.49 & 0.86 & 0.88 & 1.03 &  & 0.52 & 0.87 & 0.92 & 1.05 \\ 
$\phi _{1}$ & 0.18 & 0.68 & 0.50 & 0.86 &  & 0.98 & 0.99 & 1.39 & 1.40 \\ 
$\phi _{2}$ & --- & --- & --- & --- &  & 0.86 & 1.08 & 1.18 & 1.25 \\ 
$\psi _{1}$ & 0.17 & 0.62 & 0.47 & 0.75 &  & 0.99 & 0.99 & 1.40 & 1.40 \\ 
$\psi _{2}$ & --- & --- & --- & --- &  & 0.53 & 1.01 & 0.87 & 1.20%
\end{tabular}

\caption{Simulation Results. Bias and root mean squared errors of warped functional regression (W)
and ordinary functional regression (O) for sample size $n=50$.}

\label{tab:Simulations_1_50}

\end{table}%

\begin{table}[tbp] \centering%

\noindent 
\begin{tabular}{cccccccccc}
& \multicolumn{4}{c}{Model 1} &  & \multicolumn{4}{c}{Model 2} \\ 
& \multicolumn{2}{c}{bias} & \multicolumn{2}{c}{rmse} &  & 
\multicolumn{2}{c}{bias} & \multicolumn{2}{c}{rmse} \\ 
Param. & W & O & W & O &  & W & O & W & O \\ 
$\beta $ & 0.12 & 0.18 & 0.18 & 0.24 &  & 0.12 & 0.70 & 0.29 & 0.72 \\ 
$\mu _{x}$ & 0.10 & 0.19 & 0.27 & 0.31 &  & 0.13 & 0.19 & 0.29 & 0.30 \\ 
$\mu _{y}$ & 0.14 & 0.33 & 0.32 & 0.43 &  & 0.19 & 0.60 & 0.40 & 0.68 \\ 
$\phi _{1}$ & 0.05 & 0.05 & 0.11 & 0.13 &  & 0.07 & 0.05 & 0.19 & 0.12 \\ 
$\psi _{1}$ & 0.16 & 0.20 & 0.19 & 0.28 &  & 0.10 & 0.84 & 0.18 & 0.85 \\ 
&  &  &  &  &  &  &  &  &  \\ 
& \multicolumn{4}{c}{Model 3} &  & \multicolumn{4}{c}{Model 4} \\ 
$\beta $ & 0.38 & 1.06 & 0.83 & 1.13 &  & 0.41 & 1.26 & 0.88 & 1.31 \\ 
$\mu _{x}$ & 0.13 & 0.27 & 0.34 & 0.38 &  & 0.11 & 0.27 & 0.34 & 0.38 \\ 
$\mu _{y}$ & 0.16 & 0.38 & 0.40 & 0.49 &  & 0.18 & 0.66 & 0.45 & 0.75 \\ 
$\phi _{1}$ & 0.55 & 0.99 & 0.79 & 1.40 &  & 0.48 & 0.99 & 0.70 & 1.40 \\ 
$\phi _{2}$ & 0.22 & 1.04 & 0.46 & 1.09 &  & 0.15 & 1.04 & 0.40 & 1.09 \\ 
$\psi _{1}$ & 0.84 & 0.98 & 1.19 & 1.39 &  & 0.81 & 0.99 & 1.15 & 1.40 \\ 
$\psi _{2}$ & 0.12 & 0.92 & 0.33 & 1.13 &  & 0.16 & 0.63 & 0.34 & 1.00 \\ 
&  &  &  &  &  &  &  &  &  \\ 
& \multicolumn{4}{c}{Model 5} &  & \multicolumn{4}{c}{Model 6} \\ 
$\beta $ & 0.17 & 0.74 & 0.60 & 0.77 &  & 0.85 & 1.05 & 1.25 & 1.08 \\ 
$\mu _{x}$ & 0.43 & 0.95 & 0.69 & 1.04 &  & 0.53 & 0.95 & 0.74 & 1.03 \\ 
$\mu _{y}$ & 0.48 & 0.88 & 0.73 & 0.97 &  & 0.50 & 0.88 & 0.74 & 0.97 \\ 
$\phi _{1}$ & 0.15 & 0.76 & 0.42 & 0.87 &  & 0.99 & 0.99 & 1.40 & 1.40 \\ 
$\phi _{2}$ & --- & --- & --- & --- &  & 0.92 & 1.18 & 1.13 & 1.27 \\ 
$\psi _{1}$ & 0.16 & 0.66 & 0.40 & 0.72 &  & 0.97 & 0.99 & 1.38 & 1.40 \\ 
$\psi _{2}$ & --- & --- & --- & --- &  & 0.47 & 1.14 & 0.70 & 1.23%
\end{tabular}

\caption{Simulation Results. Bias and root mean squared errors of warped functional regression (W)
and ordinary functional regression (O) for sample size $n=100$.}

\label{tab:Simulations_1_100}

\end{table}%

We see in Tables \ref{tab:Simulations_1_50} and \ref{tab:Simulations_1_100}
that warped functional regression estimators have smaller biases than
ordinary functional regression estimators in practically all cases, which is
not surprising since the model has more parameters; for the same reason they
are going to have higher variances. The questions is whether the smaller
bias outweighs the higher variance. Root mean squared errors show that this
is indeed the case: warped regression estimators beat ordinary least squares
estimators in practically all cases. The exception is Model 6, where
covariates and responses are warped independently and the warped regression
estimator cannot fully show its advantages. However, even in this
unfavorable case the root mean squared error of the warped regression
estimator of $\beta $ is not much higher than that of the ordinary least
squares estimator, and for the other functional parameters it is actually
smaller. Therefore, from the point of view of estimation accuracy the warped
functional regression estimator is advantageous in presence of phase
variability.

\subsection{Prediction accuracy}

Another aspect of the regression problem is prediction, or the estimation of
a response function $y(t)$ for a new covariate curve $x(s)$. We compared
prediction accuracy of warped and ordinary regression estimators by
simulating data from Models 1--4 of Section \ref{subsec:Sim_estim_acc}; for
Models 5 and 6 prediction did not make much sense because covariate and
response warping functions were independent. In addition to training samples
of sizes $n=50$ and $n=100$, we generated prediction samples of size $%
n^{\ast }=100$ on equally-spaced time grids of size $\nu =20$ and measured
the prediction accuracy by the root mean squared error $\{E(\sum_{i=1}^{n^{%
\ast }}\left\Vert \mathbf{y}_{i}-\mathbf{\hat{y}}_{i}\right\Vert ^{2}/\nu
n^{\ast })\}^{1/2}$. For each model we computed the same estimators as in
Section \ref{subsec:Sim_estim_acc} and in addition ordinary linear
regression estimators with more principal components. Specifically, for the
one-dimensional models 1 and 2 we considered ordinary least squares
estimators with 1, 2 and 3 components, and for the two-dimensional models 3
and 4 we considered estimators with 2, 3 and 4 components.

\begin{table}[tbp] \centering%

\begin{tabular}{ccccc}
& \multicolumn{2}{c}{Model 1} & \multicolumn{2}{c}{Model 2} \\ 
Estim. & $n=50$ & $n=100$ & $n=50$ & $n=100$ \\ 
W-1 & 0.14 & 0.13 & 0.15 & 0.14 \\ 
O-1 & 0.19 & 0.19 & 0.20 & 0.20 \\ 
O-4 & 0.14 & 0.13 & 0.15 & 0.15 \\ 
O-9 & 0.14 & 0.13 & 0.15 & 0.15 \\ 
&  &  &  &  \\ 
& \multicolumn{2}{c}{Model 3} & \multicolumn{2}{c}{Model 4} \\ 
W-4 & 0.20 & 0.19 & 0.21 & 0.20 \\ 
O-4 & 0.21 & 0.20 & 0.23 & 0.23 \\ 
O-9 & 0.17 & 0.17 & 0.20 & 0.19 \\ 
O-16 & 0.17 & 0.16 & 0.19 & 0.18%
\end{tabular}

\caption{Simulation Results. Prediction errors for new responses using 
warped functional regression (W) and ordinary functional regression (O).}

\label{tab:Simulations_2}

\end{table}%

Table \ref{tab:Simulations_2} shows the results. The table indicates the
overall dimension of the estimators: for example, O-9 is the ordinary
regression estimator based on 3 principal components for covariates and
responses, which has overall dimension 9. Prediction errors of ordinary
linear regression estimators will decrease as the number of principal
components increases, and eventually they will be smaller than prediction
errors of warped regression estimators of fixed dimension. The point is that
given comparable prediction errors, a low-dimensional warped regression
model that neatly separates the two sources of variability will be
preferable to a higher-dimensional ordinary linear model that confounds them.

We see that, generally speaking, the ordinary linear regression estimator
needs an additional principal component to attain a comparable or smaller
prediction error than the warped regression estimator, although sometimes a
strictly smaller prediction error is not attained, as in Models 1 and 2. For
Models 3 and 4 the ordinary least squares estimator does attain smaller
prediction errors, but in order to attain an error that is only 10\% smaller
it needs to use four times as many parameters as the warped regression
model, which makes it extremely impractical from the point of view of
interpretability. Interpretability issues cannot be directly gleaned from
Table \ref{tab:Simulations_2} or other simulation summaries because they are
graphical in nature, so we are going to study them by example in \S\ \ref%
{sec:Example}.

\subsection{Asymptotic accuracy}

We also studied by simulation the finite-sample adequacy of the asymptotic
results of \S\ \ref{sec:Inference}, particularly for hypothesis testing. We
simulated data from Model 1 with $\mathbf{A}=\mathbf{0}$, and also from a
similar model that uses equally-spaced time grids of size 15 instead of the
random time grids of Model 1. Two sample sizes were considered in each case, 
$n=50$ and $n=200$. Each scenario was replicated 500 times.

The warped regression estimator was computed using the same specifications
as above. The covariance matrix of $\mathrm{vec}(\mathbf{\hat{A}}^{T})$ was
estimated by the asymptotic formulas of \S\ \ref{sec:Inference} and by
bootstrap, using 50 bootstrap samples. The `true' covariance matrix of $%
\mathrm{vec}(\mathbf{\hat{A}}^{T})$ was computed as the sample covariance of
the 500 replicated estimators. Since we are interested in testing, we
computed tail probabilities of $Q=\mathrm{vec}(\mathbf{\hat{A}}^{T})^{T}%
\mathbf{\hat{\Sigma}}^{-1}\mathrm{vec}(\mathbf{\hat{A}}^{T})$, where $%
\mathbf{\hat{\Sigma}}$ is the respective covariance estimator of $\mathrm{vec%
}(\mathbf{\hat{A}}^{T})$, and of $Z_{1j}=\hat{a}_{1j}/\widehat{\mathrm{sd}}(%
\hat{a}_{1j})$ for $j=1,2$. Specifically, we report $P\left( Q\geq
7.78\right) $ and $P\left( \left\vert Z_{1j}\right\vert \geq 1.645\right) $
for $j=1,2$, which should be close to $0.10$.

\begin{table}[tbp] \centering%

\begin{tabular}{llccccccc}
&  & \multicolumn{7}{c}{Random grids} \\ 
&  & \multicolumn{3}{c}{$n=50$} &  & \multicolumn{3}{c}{$n=200$} \\ 
&  & $Q$ & $Z_{11}$ & $Z_{12}$ &  & $Q$ & $Z_{11}$ & $Z_{12}$ \\ 
True variance &  & 0.09 & 0.08 & 0.08 &  & 0.10 & 0.10 & 0.09 \\ 
Asymptotic &  & 0.34 & 0.24 & 0.21 &  & 0.33 & 0.21 & 0.20 \\ 
Bootstrap &  & 0.25 & 0.16 & 0.13 &  & 0.25 & 0.14 & 0.11 \\ 
&  & \multicolumn{1}{l}{} & \multicolumn{1}{l}{} & \multicolumn{1}{l}{} & 
\multicolumn{1}{l}{} & \multicolumn{1}{l}{} & \multicolumn{1}{l}{} & 
\multicolumn{1}{l}{} \\ 
&  & \multicolumn{7}{c}{Equally spaced grids} \\ 
True variance &  & 0.11 & 0.08 & 0.08 &  & 0.10 & 0.10 & 0.07 \\ 
Asymptotic &  & 0.36 & 0.20 & 0.23 &  & 0.27 & 0.14 & 0.26 \\ 
Bootstrap &  & 0.33 & 0.18 & 0.21 &  & 0.29 & 0.11 & 0.23%
\end{tabular}

\caption{Simulation Results. Tail probabilities of test statistics, true
value is 0.10.}

\label{tab:Simulations_3}

\end{table}%

Table \ref{tab:Simulations_3} shows the results. There are two aspects of
the asymptotics that we are trying to assess: the adequacy of the normal
approximation and the adequacy of the variance estimators. The first aspect
can be best assessed using the true variance in the test statistics, so the
variance estimation error is not a confounding factor. In this regard we see
in Table \ref{tab:Simulations_3} that the asymptotic approximation is good
even for $n=50$, both for the global $Q$-test and for the marginal $Z$%
-tests. In the more realistic cases where the variance is estimated, we see
that bootstrap variance estimators generally work better than the
asymptotic-variance formula; although both underestimate the true variances,
bootstrap tends to underestimate them less, especially for random time grids.

\section{\label{sec:Example}Application: Modeling Ground-Level Ozone
Concentration}

Ground-level ozone is an air pollutant known to cause serious health
problems. Unlike other pollutants, ozone is not emitted directly into the
air but is a result of complex chemical reactions in the atmosphere that
include, among other factors, volatile organic compounds and oxides of
nitrogen. Oxides of nitrogen are emitted by combustion engines, power plants
and other industrial sources. The modeling of ground-level ozone formation
has been an active topic of air-quality studies for many years.

In this article we will use data from the California Environmental
Protection Agency online database. Hourly concentration of pollutants at
many locations in California are available for the years 1980--2009. We will
analyze trajectories of oxides of nitrogen (NOx) and ozone (O3) in the city
of Sacramento (site 3011 in the database) in the Summer of 2005. We omit
weekends and holidays because NOx and O3 levels are substantially lower and
follow different patterns. We also removed some outlying trajectories, so
the final sample consisted of 52 days between June 6 and August 26, shown in
Figure \ref{fig:Sample_curves}.

Both NOx and O3 trajectories follow simple regular patterns. NOx curves tend
to peak around 7am, and O3 curves around 2pm. Therefore we fitted warped
regression models with single warping knots, trying several values of $\tau
_{x0}$ and $\tau _{y0}$ around 7am and 2pm respectively. The results were
similar in all cases; the estimators reported here correspond to $\tau
_{x0}=7$ and $\tau _{y0}=14$. As basis functions we used cubic B-splines
with 7 equally spaced knots, one knot every 3 hours; we also tried 10 knots
but the results were not substantially different. Three warped regression
models were fitted: \emph{(i)} a model with\emph{\ }one principal component
for $x$ and one for $y$, \emph{(ii)} a model with\emph{\ }two principal
components for $x$ and one for $y$, and \emph{(iii)} a model with one
principal component for $x$ and two for $y$. The log-likelihood values were
44.44, 45.21 and 52.04, respectively. The second model did not seem to
represent much of an improvement over the first one, so we discarded it. For
models \emph{(i)} and \emph{(iii)} the estimated regression coefficients and
the bootstrap standard deviations, based on 200 resamples, were 
\begin{eqnarray*}
\mathbf{\hat{A}} &=&\left[ 
\begin{array}{cc}
0.73 & 0.09 \\ 
0.19 & 0.44%
\end{array}%
\right] ,\ \ \mathrm{std}(\mathbf{\hat{A})}=\left[ 
\begin{array}{cc}
0.07 & 0.02 \\ 
0.08 & 0.06%
\end{array}%
\right] , \\
\mathbf{\hat{A}} &=&\left[ 
\begin{array}{cc}
0.36 & 0.12 \\ 
0.01 & 0.02 \\ 
0.18 & 0.54%
\end{array}%
\right] ,\ \ \mathrm{std}(\mathbf{\hat{A})}=\left[ 
\begin{array}{cc}
0.08 & 0.06 \\ 
0.04 & 0.10 \\ 
0.06 & 0.11%
\end{array}%
\right] .
\end{eqnarray*}

\FRAME{ftbpFU}{5.7865in}{5.9153in}{0pt}{\Qcb{Ozone Example. Warped
Functional Regression fit. (a) Log-NOx mean (solid line), and mean plus
(dashed line) and minus (dotted line) the principal component; (b) same as
(a) for the square root of O3; (c) covariate versus response pc-scores; (d)
covariate peak versus response pc-score; (e) covariate pc-score versus
response peak; (f) covariate versus response peaks.}}{\Qlb{fig:WFR_fit}}{%
wfr_fit.eps}{\special{language "Scientific Word";type
"GRAPHIC";maintain-aspect-ratio TRUE;display "USEDEF";valid_file "F";width
5.7865in;height 5.9153in;depth 0pt;original-width 9.4438in;original-height
6.1557in;cropleft "0";croptop "1";cropright "1";cropbottom "0";filename
'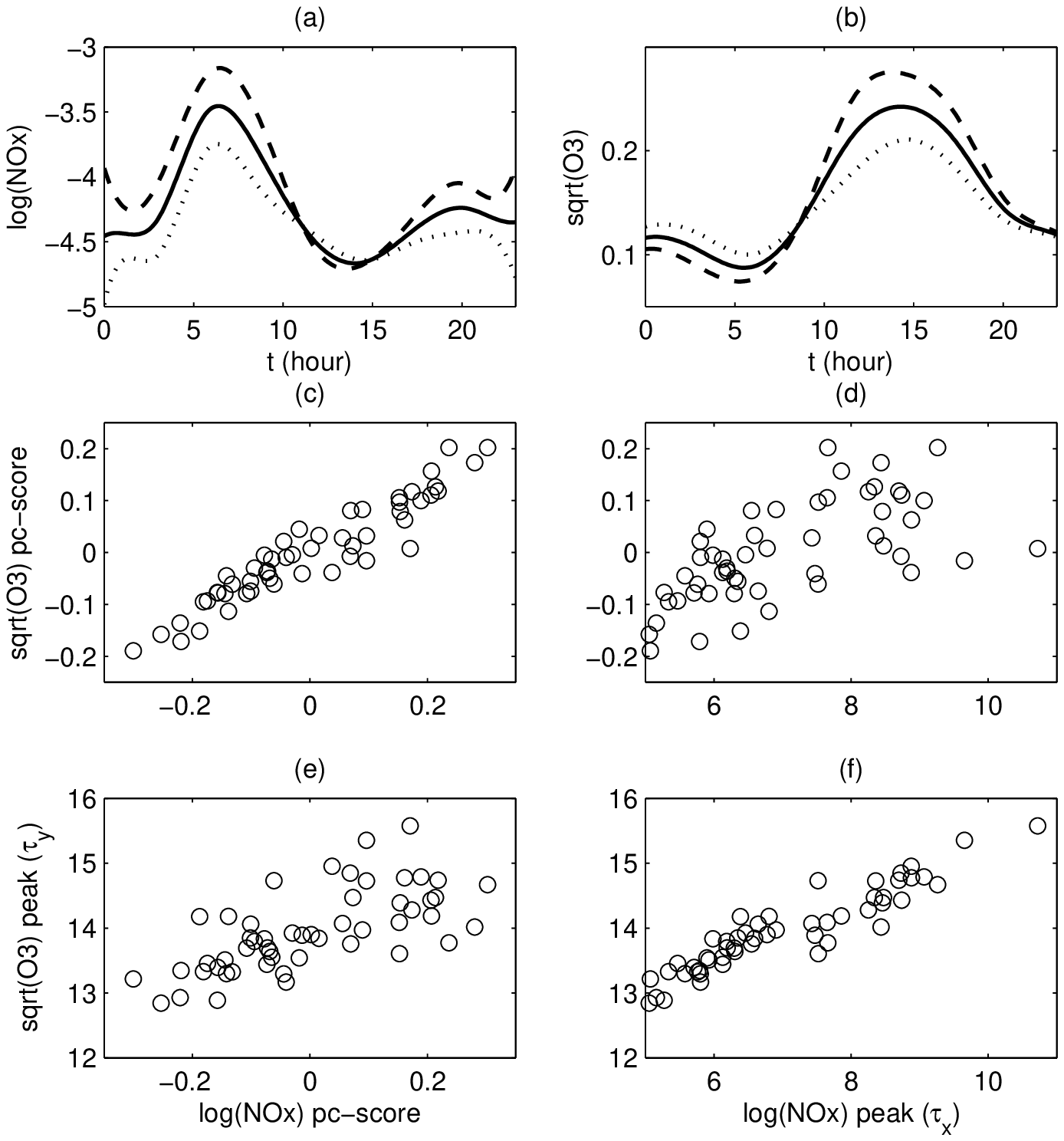';file-properties "XNPEU";}}

For model (iii) the coefficients of the second principal component of the
response, $\hat{a}_{21}$ and $\hat{a}_{22}$, are not significant, while for
model (i) all coefficients are significant even allowing for underestimation
of the standard deviations, with the possible exception of $\hat{a}_{21}$
which is a borderline case. For this reason we prefer (i) as our final
model. To interpret the principal components, Figure \ref{fig:WFR_fit}(a)
shows $\hat{\mu}_{x}$ and $\hat{\mu}_{x}\pm c_{1}\hat{\phi}_{1}$ for some
constant $c_{1}$, and Figure \ref{fig:WFR_fit}(b) shows $\hat{\mu}_{y}$ and $%
\hat{\mu}_{y}\pm c_{2}\hat{\psi}_{1}$ for another constant $c_{2}$. Both
principal components are shape components: curves with positive scores tend
to have sharper features than the mean while curves with negative scores
tend to have flatter features than the mean. The fact that the diagonal
coefficients of $\mathbf{\hat{A}}$ are positive indicates that the component
scores $\hat{u}_{i}$ and $\hat{v}_{i}$ are positively correlated, as Figure %
\ref{fig:WFR_fit}(c) shows, and the warping landmarks $\hat{\tau}_{xi}$ and $%
\hat{\tau}_{yi}$, which can roughly be interpreted as peak locations, are
also positively correlated, as Figure \ref{fig:WFR_fit}(f) shows. Amplitude
and warping factors are also positively cross-correlated, since the
off-diagonal elements of $\mathbf{\hat{A}}$ are also positive. In particular 
$\hat{a}_{12}$ is highly significant, so late NOx peaks tend to be
associated with high peaks of O3 and vice-versa, as Figure \ref{fig:WFR_fit}%
(d) shows.

\FRAME{ftbpFU}{5.5365in}{5.9162in}{0pt}{\Qcb{Ozone Example. Ordinary
Functional Regression fit. (a,c,e) Mean (solid line), and mean plus (dashed
line) and minus (dotted line) the first [(a)], second [(c)] and third [(e)]
principal components of explanatory curves; (b,d,f) same as (a,c,e),
respectively, for response curves.}}{\Qlb{fig:FLR_fit}}{flr_fit.eps}{\special%
{language "Scientific Word";type "GRAPHIC";maintain-aspect-ratio
TRUE;display "USEDEF";valid_file "F";width 5.5365in;height 5.9162in;depth
0pt;original-width 6.1203in;original-height 6.5414in;cropleft "0";croptop
"1";cropright "1";cropbottom "0";filename '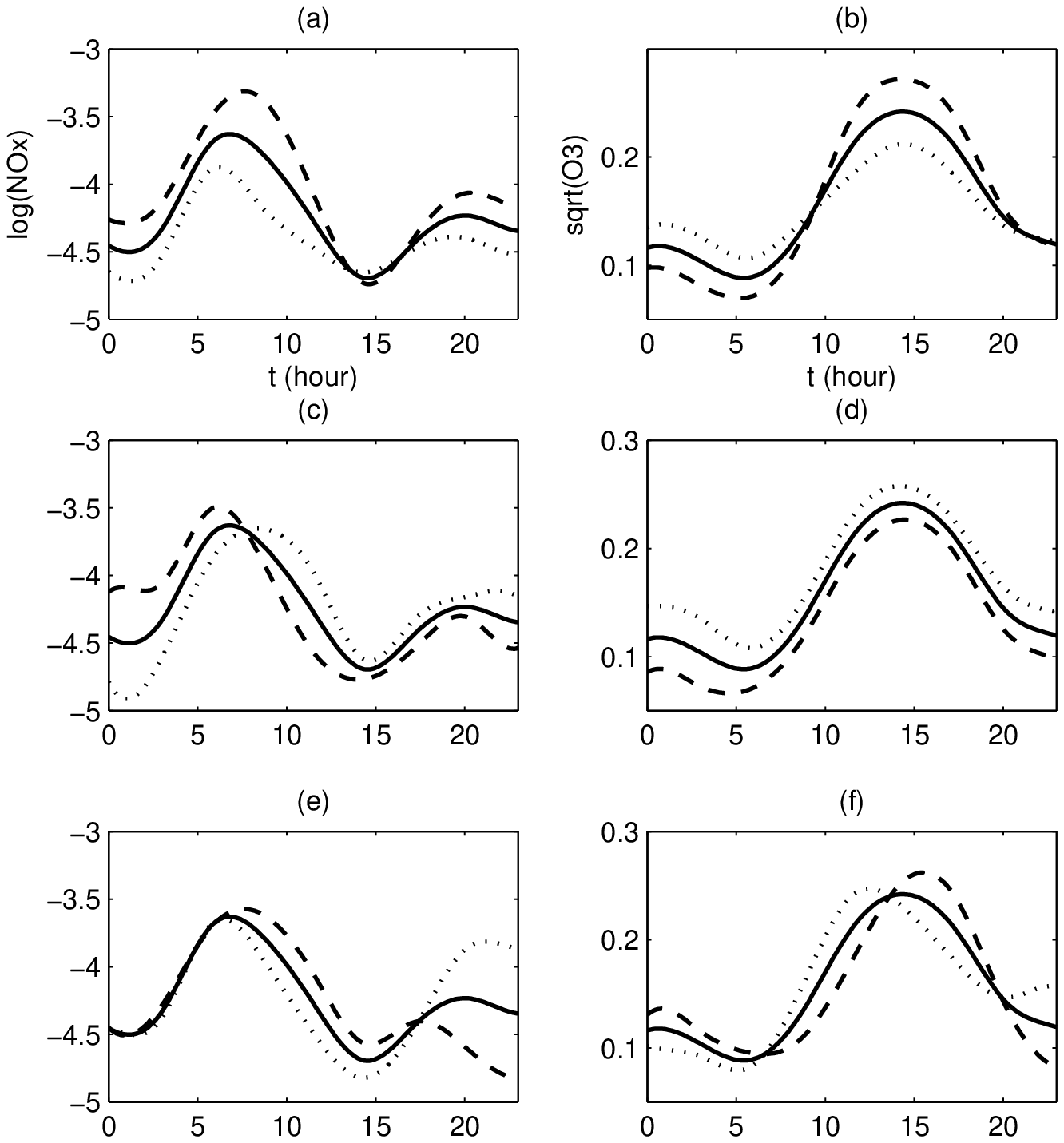';file-properties
"XNPEU";}}

An ordinary functional regression fit is shown in Figure \ref{fig:FLR_fit};
the plot shows $\hat{\mu}_{x}$, $\hat{\mu}_{y}$, $\hat{\mu}_{x}\pm c_{1}\hat{%
\phi}_{j}$ and $\hat{\mu}_{y}\pm c_{2}\hat{\psi}_{j}$ for a three-component
model, or overall dimension 9. A two-component model, of overall dimension 4
and thus comparable to the warped regression model, would correspond to the
upper four panels of Figure \ref{fig:FLR_fit}. Time variability in the
explanatory curves is explained by the second $x$-component (Figure \ref%
{fig:FLR_fit}(c)), but phase variability in the response curves is not
accounted for until the third component (Figure \ref{fig:FLR_fit}(f)), so it
really takes a 9-dimensional ordinary regression model to explain the
phase-variability features that a 4-dimensional warped model would explain.
And the predominantly time-related principal components, Figure \ref%
{fig:FLR_fit}(c,f), are also associated with some kinds of amplitude
variability. Likewise, principal components that are predominantly
amplitude-related, like the first $x$-component, Figure \ref{fig:FLR_fit}%
(a), are somewhat influenced by time variability. This blurring of the
components is avoided by warped functional regression, which neatly
separates the sources of variability and offers not only a more easily
interpretable model but also a lower-dimensional one.

\section*{Acknowledgement}

This research was partially supported by a grant from the National Science
Foundation.

\section*{Supplementary material}

Supplementary material available online includes a more thorough discussion
of model identifiability, the derivation of the EM algorithm for estimation,
detailed derivation of formulae involved in the asymptotic distribution of
the estimator, and a detailed treatment of monotone Hermite splines.

\section*{Appendix}

\subsection{\label{app:Hermite}Monotone Hermite splines}

In this section we explain how the warping functions $\omega _{i}(s)$ are
constructed; the $\zeta _{i}(t)$s are constructed in a similar way. Let $%
\mathcal{S}=[a,b]$ and $a<\tau _{01}<\cdots <\tau _{0r}<b$ be a sequence of $%
r$ knots in $\mathcal{S}$. Define the basis functions $\{\alpha _{j}(s;%
\mathbf{\tau }_{0})\}$ and $\{\beta _{j}(s;\mathbf{\tau }_{0})\}$ as
follows: let $h_{00}(s)=(1+2s)(1-s)^{2}$ and $h_{10}(s)=s(1-s)^{2}$; then 
\[
\alpha _{0}(s;\mathbf{\tau }_{0})=\left\{ 
\begin{array}{lll}
0 &  & \text{if }s<a\text{ or }s>\tau _{01} \\ 
h_{00}\left( \frac{s-a}{\tau _{01}-a}\right) &  & \text{if }a\leq s\leq \tau
_{01},%
\end{array}%
\right. 
\]%
\[
\alpha _{j}(s;\mathbf{\tau }_{0})=\left\{ 
\begin{array}{lll}
0 &  & \text{if }s<\tau _{0,j-1}\text{ or }s>\tau _{0,j+1} \\ 
h_{00}\left( \frac{\tau _{0j}-s}{\tau _{0j}-\tau _{0,j-1}}\right) &  & \text{%
if }\tau _{0,j-1}\leq s\leq \tau _{0j} \\ 
h_{00}\left( \frac{s-\tau _{0j}}{\tau _{0,j+1}-\tau _{0j}}\right) &  & \text{%
if }\tau _{0j}\leq s\leq \tau _{0,j+1}%
\end{array}%
\right. 
\]%
for $j=1,\ldots ,r$, 
\[
\alpha _{r+1}(s;\mathbf{\tau }_{0})=\left\{ 
\begin{array}{lll}
0 &  & \text{if }s<\tau _{0r}\text{ or }s>b \\ 
h_{00}\left( \frac{b-s}{b-\tau _{0r}}\right) &  & \text{if }\tau _{0r}\leq
s\leq b,%
\end{array}%
\right. 
\]%
\[
\beta _{0}(s;\mathbf{\tau }_{0})=\left\{ 
\begin{array}{lll}
0 &  & \text{if }s<a\text{ or }s>\tau _{01} \\ 
(\tau _{01}-a)h_{10}\left( \frac{s-a}{\tau _{01}-a}\right) &  & \text{if }%
a\leq s\leq \tau _{01},%
\end{array}%
\right. 
\]%
\[
\beta _{j}(s;\mathbf{\tau }_{0})=\left\{ 
\begin{array}{lll}
0 &  & \text{if }s<\tau _{0,j-1}\text{ or }s>\tau _{0,j+1} \\ 
-(\tau _{0j}-\tau _{0,j-1})h_{10}\left( \frac{\tau _{0j}-s}{\tau _{0j}-\tau
_{0,j-1}}\right) &  & \text{if }\tau _{0,j-1}\leq s\leq \tau _{0,j} \\ 
(\tau _{0,j+1}-\tau _{0,j})h_{10}\left( \frac{s-\tau _{0,j}}{\tau
_{0,j+1}-\tau _{0,j}}\right) &  & \text{if }\tau _{0,j}\leq s\leq \tau
_{0,j+1}%
\end{array}%
\right. 
\]%
for $j=1,\ldots ,r$, and 
\[
\beta _{r+1}(s;\mathbf{\tau }_{0})=\left\{ 
\begin{array}{lll}
0 &  & \text{if }s<\tau _{0r}\text{ or }s>b \\ 
-(b-\tau _{0r})h_{10}\left( \frac{b-s}{b-\tau _{0r}}\right) &  & \text{if }%
\tau _{0r}\leq s\leq b.%
\end{array}%
\right. 
\]%
The function 
\begin{equation}
\omega _{i}(s)=\sum_{j=0}^{r+1}\tau _{ij}\alpha _{j}(s;\mathbf{\tau }%
_{0})+\sum_{j=0}^{r+1}d_{ij}\beta _{j}(s;\mathbf{\tau }_{0}),
\label{eq:w_lin_exp}
\end{equation}%
where $\tau _{i0}=a$ and $\tau _{i,r+1}=b$, is a differentiable
piecewise-cubic function that satisfies $\omega _{i}(\tau _{0j})=\tau _{ij}$
and $\omega _{i}^{\prime }(\tau _{0j})=d_{ij}$ for $j=1,\ldots ,r$. Thus the 
$\tau _{ij}$s play the dual role of basis coefficients and values of $\omega
_{i}(s)$ at the knots. For (\ref{eq:w_lin_exp}) to be strictly monotone
increasing the $d_{ij}$s must satisfy certain necessary and sufficient
conditions given in Fritsch \& Carlson (1980). For situations like ours
where no particular values of the $d_{ij}$s are specified, Fritsch \&
Carlson provide a simple algorithm to compute, from given $\tau _{ij}$s,
values of the $d_{ij}$s that satisfy the monotonicity constraints. This
algorithm is given in the Supplementary Material. Since the algorithm is
deterministic, the $d_{ij}$s are functions of the $\tau _{ij}$s and then (%
\ref{eq:w_lin_exp}) is entirely parameterized by $\mathbf{\tau }_{i}=(\tau
_{i1},\ldots ,\tau _{ir})$, thus forming an $r$-dimensional space.

The Jupp transform (Jupp, 1978) is defined as 
\[
\theta _{ij}=\log \left( \frac{\tau _{i,j+1}-\tau _{ij}}{\tau _{ij}-\tau
_{i,j-1}}\right) ,\ \ j=1,\ldots ,r, 
\]%
with inverse given by 
\[
\tau _{ij}=a+(b-a)\cdot \frac{\sum_{k=1}^{j}\exp (\theta _{i1}+\cdots
+\theta _{ik})}{\{1+\sum_{k=1}^{r}\exp (\theta _{i1}+\cdots +\theta _{ik})\}}%
,\ \ j=1,\ldots ,r. 
\]%
Note that for any $r$-dimensional unconstrained vector $\mathbf{\theta }$
the inverse Jupp transform yields a vector $\mathbf{\tau }$ of strictly
increasing knots in $(a,b)$. In particular, for $\mathbf{\theta }=\mathbf{0}$
the corresponding $\mathbf{\tau }$ is a sequence of $r$ equally spaced knots
in $(a,b)$.

\subsection{\label{app:Likelihood}Likelihood function}

Under the distributional assumptions in Section \ref{sec:Estimation}, the
likelihood function is derived as follows. The joint density function of the
data vectors $(\mathbf{x}_{i},\mathbf{y}_{i})$ and the latent random effects 
$(\mathbf{w}_{i},\mathbf{z}_{i})$ can be factorized as 
\begin{eqnarray*}
f(\mathbf{x}_{i},\mathbf{y}_{i},\mathbf{w}_{i},\mathbf{z}_{i}) &=&f(\mathbf{x%
}_{i},\mathbf{y}_{i}|\mathbf{w}_{i},\mathbf{z}_{i})f(\mathbf{z}_{i}|\mathbf{w%
}_{i})f(\mathbf{w}_{i}) \\
&=&f(\mathbf{x}_{i}|\mathbf{w}_{i})f(\mathbf{y}_{i}|\mathbf{z}_{i})f(\mathbf{%
z}_{i}|\mathbf{w}_{i})f(\mathbf{w}_{i}),
\end{eqnarray*}%
since $\mathbf{y}_{i}$ depends on $\mathbf{w}_{i}$ only through $\mathbf{z}%
_{i}$, according to (\ref{eq:Reg-model}). Clearly $\mathbf{w}_{i}\sim N(%
\mathbf{\mu }_{w},\mathbf{\Sigma }_{w})$ and $\mathbf{z}_{i}|\mathbf{w}%
_{i}\sim N\{\mathbf{\mu }_{z}+\mathbf{A}\left( \mathbf{w}_{i}-\mathbf{\mu }%
_{w}\right) ,\mathbf{\Sigma }_{e}\}$. The conditional distributions $\mathbf{%
x}_{i}|\mathbf{w}_{i}$ and $\mathbf{y}_{i}|\mathbf{z}_{i}$ are derived as
follows. Given $\mathbf{w}_{i}=(\mathbf{u}_{i}^{T},\mathbf{\theta }%
_{xi}^{T})^{T}$ and $\mathbf{z}_{i}=(\mathbf{v}_{i}^{T},\mathbf{\theta }%
_{yi}^{T})^{T}$, the values of $\mathbf{\theta }_{xi}$ and $\mathbf{\theta }%
_{yi}$ determine the warping functions $\omega _{i}(s)$ and $\zeta _{i}(t)$
and consequently two warped time grids $s_{ij}^{\ast }=\omega
_{i}^{-1}(s_{ij})$, $j=1,\ldots ,\nu _{1i}$, and $t_{ij}^{\ast }=\zeta
_{i}^{-1}(t_{ij})$, $j=1,\ldots ,\nu _{2i}$. Let $\mathbf{B}_{xi}^{\ast }\in 
\mathbb{R}^{\nu _{1i}\times q_{1}}$ and $\mathbf{B}_{yi}^{\ast }\in \mathbb{R%
}^{\nu _{2i}\times q_{2}}$ be the B-spline bases evaluated at the warped
time grids, that is $\mathbf{B}_{xi,jk}^{\ast }=b_{xk}(s_{ij}^{\ast })$ and $%
\mathbf{B}_{yi,jk}^{\ast }=b_{yk}(t_{ij}^{\ast })$. Then, in view of model
specifications (\ref{eq:raw_data_x})--(\ref{eq:KL-model-y}) we have $\mathbf{%
x}_{i}|\mathbf{w}_{i}\sim N(\mathbf{B}_{xi}^{\ast }\mathbf{m}_{x}+\mathbf{B}%
_{xi}^{\ast }\mathbf{Cu}_{i},\sigma _{\varepsilon }^{2}\mathbf{I}_{\nu
_{1i}})$ and $\mathbf{y}_{i}|\mathbf{z}_{i}\sim N(\mathbf{B}_{yi}^{\ast }%
\mathbf{m}_{y}+\mathbf{B}_{yi}^{\ast }\mathbf{Dv}_{i},\sigma _{\eta }^{2}%
\mathbf{I}_{\nu _{2i}})$. The maximum likelihood estimators maximize 
\begin{equation}
\ell (\mathbf{A},\mathbf{\Sigma }_{e},\mathbf{\Sigma }_{w},\mathbf{m}_{x},%
\mathbf{m}_{y},\mathbf{C},\mathbf{D},\sigma _{\varepsilon }^{2},\sigma
_{\eta }^{2})=\sum_{i=1}^{n}\log \iint f(\mathbf{x}_{i},\mathbf{y}_{i},%
\mathbf{w},\mathbf{z})\ \mathrm{d}\mathbf{w\ }\mathrm{d}\mathbf{z}
\label{eq:loglik}
\end{equation}%
but the integrals in (\ref{eq:loglik}) do not have closed forms so we use
the EM algorithm to find the optimum, treating the random effects $(\mathbf{w%
}_{i},\mathbf{z}_{i})$ as missing data. Most of the updating equations of
the EM algorithm are easy to derive but the restrictions on the parameters $%
\mathbf{C}$, $\mathbf{D}$, and $\mathbf{A}$ pose some difficulties. This is
discussed in detail in the Supplementary Material.

\subsection*{\label{app:Theo_proof}Proof of Theorem \protect\ref{theo:Asymp}}

This proof is a direct application of Theorem 4.4 of Geyer (1994); note that
Theorem 5.2 of Geyer (1994), which pertains to consistent local minimizers
instead of global minimizers, can also be applied because our $T_{C}(\mathbf{%
\zeta }_{0})$ satisfies the stronger condition of being Clarke-regular
(Rockafellar \& Wets, 1998, ch.~6.B). Following Geyer's notation, let $F(%
\mathbf{\zeta })=E\{-\log f(\mathbf{x,y};\mathbf{\zeta })\}$ and $F_{n}(%
\mathbf{\zeta })=-(1/n)\sum_{i=1}^{n}\log f(\mathbf{x}_{i}\mathbf{,y}_{i};%
\mathbf{\zeta })$. Then $\mathbf{\hat{\zeta}}_{n}=\arg \min_{\mathbf{\zeta }%
\in C}F_{n}(\mathbf{\zeta })$ and $\mathbf{\zeta }_{0}=\arg \min_{\mathbf{%
\zeta }\in C}F(\mathbf{\zeta })$. Assumption A of Geyer (1994) is that 
\begin{equation}
F(\mathbf{\zeta })=F(\mathbf{\zeta }_{0})+\frac{1}{2}(\mathbf{\zeta }-%
\mathbf{\zeta }_{0})^{T}\mathbf{V}(\mathbf{\zeta }-\mathbf{\zeta }%
_{0})+o(\Vert \mathbf{\zeta }-\mathbf{\zeta }_{0}\Vert ),  \label{eq:exp_F}
\end{equation}%
with $\mathbf{V}=\nabla ^{2}F(\mathbf{\zeta }_{0})$ positive definite. This
is satisfied in our case because $\nabla F(\mathbf{\zeta }_{0})=-E\{\nabla
\log f(\mathbf{x,y};\mathbf{\zeta }_{0})\}=\mathbf{0}$ and $\nabla ^{2}F(%
\mathbf{\zeta }_{0})=E\{\mathbf{U(x,y)U(x,y)}^{T}\}$. To see that the latter
is positive definite, note that for $\mathbf{\zeta }$ as in (\ref{eq:theta})
we have 
\begin{eqnarray*}
\mathbf{U(x,y)}^{T}\mathbf{\zeta } &=&\mathrm{tr}\{\mathbf{\Sigma }%
_{e,0}^{-1}\mathbf{N(x,y)}^{T}\mathbf{A}^{T}\}-\mathrm{tr}\{\mathbf{\Sigma }%
_{e,0}^{-1}\mathbf{A}_{0}\mathbf{M(x,y)A}^{T}\} \\
&&-\frac{1}{2}\mathrm{tr}\{\mathbf{\Sigma }_{w,0}^{-1}\mathbf{\Sigma }_{w}-%
\mathbf{\Sigma }_{w,0}^{-1}\mathbf{M(x,y)\Sigma }_{w,0}^{-1}\mathbf{\Sigma }%
_{w}\} \\
&=&E\{(\mathbf{w}-\mathbf{\mu }_{w})^{T}\mathbf{A}^{T}\mathbf{\Sigma }%
_{e,0}^{-1}\mathbf{e|(x},\mathbf{y})\} \\
&&-\frac{1}{2}\mathrm{tr}\{\mathbf{\Sigma }_{w,0}^{-1}\mathbf{\Sigma }_{w}\}+%
\frac{1}{2}E\{(\mathbf{w}-\mathbf{\mu }_{w})^{T}\mathbf{\Sigma }_{w,0}^{-1}%
\mathbf{\Sigma }_{w}\mathbf{\Sigma }_{w,0}^{-1}(\mathbf{w}-\mathbf{\mu }_{w})%
\mathbf{|(x},\mathbf{y})\},
\end{eqnarray*}%
where $\mathbf{e}=\mathbf{z}-\mathbf{\mu }_{z}+\mathbf{A}_{0}\left( \mathbf{w%
}-\mathbf{\mu }_{w}\right) $, then $\mathbf{\zeta }^{T}\mathbf{V\zeta }=E[\{%
\mathbf{U(x,y)}^{T}\mathbf{\zeta \}}^{2}]\geq 0$ and it is equal to zero
only if $\mathbf{U(x,y)}^{T}\mathbf{\zeta }=0$ with probability one, which
can only happen if $\mathbf{\zeta =0}$.

Assumption B of Geyer, in our case, is that 
\[
-\log f(\mathbf{x,y};\mathbf{\zeta })=-\log f(\mathbf{x,y};\mathbf{\zeta }%
_{0})+(\mathbf{\zeta }-\mathbf{\zeta }_{0})^{T}\mathbf{D}(\mathbf{x,y}%
)+\Vert \mathbf{\zeta }-\mathbf{\zeta }_{0}\Vert r(\mathbf{x,y,\zeta }) 
\]%
for some $\mathbf{D}(\mathbf{x,y})$ such that the remainder $r(\mathbf{%
x,y,\zeta })$ is stochastically equicontinuous. This is satisfied by $%
\mathbf{D}(\mathbf{x,y})=-\nabla \log f(\mathbf{x,y};\mathbf{\zeta }_{0})$;
the fact that $r(\mathbf{x,y,\zeta })$ is stochastically equicontinuous
follows from Pollard (1984, pp.~150--152). Clearly $\mathbf{D}(\mathbf{x,y})$
satisfies a Central Limit Theorem with asymptotic covariance matrix $\mathbf{%
A}$ that in this case is equal to $\mathbf{V}$, so Assumption C of Geyer is
also satisfied. Then Theorem 4.4 of Geyer can be applied. It states that the
asymptotic distribution of $\sqrt{n}(\mathbf{\hat{\zeta}}_{n}-\mathbf{\zeta }%
_{0})$ is the same as the distribution of $\mathbf{\hat{\delta}}(\mathbf{Z})$%
, the minimizer of 
\[
q_{\mathbf{Z}}(\mathbf{\delta })=\mathbf{\delta }^{T}\mathbf{Z}+\frac{1}{2}%
\mathbf{\delta }^{T}\mathbf{V\delta } 
\]%
over $\mathbf{\delta }\in T_{C}(\mathbf{\zeta }_{0})$, where $\mathbf{Z}\sim
N(\mathbf{0,A})$.

In our case $\mathbf{\hat{\delta}}(\mathbf{Z})$ can be obtained in closed
form, due to the simplicity of $T_{C}(\mathbf{\zeta }_{0})$. Concretely, $%
T_{C}(\mathbf{\zeta }_{0})$ is the space of $\mathbf{\delta }$s such that $%
\mathbf{B\delta =0}$. Let $\mathbf{\Omega }=[\mathbf{\Xi }^{\ast },\mathbf{%
\Xi }]$ be a $d\times d$ orthogonal matrix whose first $m$ columns $\mathbf{%
\Xi }^{\ast }$ span the space generated by the rows of $\mathbf{B}$ and
whose last $d-m$ columns $\mathbf{\Xi }$ are orthogonal to the rows of $%
\mathbf{B}$. Then $\mathbf{\delta }\in $ $T_{C}(\mathbf{\zeta }_{0})$ if and
only if $\mathbf{\delta =\Omega \beta }$ with $\beta _{1}=\cdots =\beta
_{m}=0$; that is, $\mathbf{\delta =\Xi \beta }_{2}$ with $\mathbf{\beta }%
_{2} $ the subvector containing the last $d-m$ coordinates of $\mathbf{\beta 
}$. Then for $\mathbf{\delta }\in $ $T_{C}(\mathbf{\zeta }_{0})$ we can
write 
\begin{eqnarray*}
q_{\mathbf{Z}}(\mathbf{\delta }) &=&\mathbf{\beta }^{T}\mathbf{\Omega }^{T}%
\mathbf{Z}+\frac{1}{2}\mathbf{\beta }^{T}\mathbf{\Omega }^{T}\mathbf{V\Omega
\beta } \\
&=&\mathbf{\beta }_{2}^{T}\mathbf{\Xi }^{T}\mathbf{Z}+\frac{1}{2}\mathbf{%
\beta }_{2}^{T}\mathbf{\Xi }^{T}\mathbf{V\Xi \beta }_{2},
\end{eqnarray*}%
which is clearly minimized by $\mathbf{\hat{\beta}}_{2}=(\mathbf{\Xi }^{T}%
\mathbf{V\Xi })^{-1}\mathbf{\Xi }^{T}\mathbf{Z}$. Therefore $\mathbf{\hat{%
\delta}}(\mathbf{Z})=\mathbf{\Xi }(\mathbf{\Xi }^{T}\mathbf{V\mathbf{\Xi }}%
)^{-1}\mathbf{\Xi }^{T}\mathbf{Z}$, and since $\mathbf{A=V}$, the result of
the theorem follows.

\section*{References}

\begin{description}
\item Ash, R.B. \& Gardner, M.F. (1975). \emph{Topics in stochastic processes%
}. New York: Academic Press.

\item Brumback, L.C.~\& Lindstrom, M.J. (2004). Self modeling with flexible,
random time transformations. \emph{Biometrics }\textbf{60} 461--470.

\item Cai, T. \& Hall, P. (2006). Prediction in functional linear
regression. \emph{The Annals of Statistics} \textbf{34} 2159--2179.

\item Crambes, C., Kneip, A., \& Sarda, P. (2009). Smoothing splines
estimators for functional linear regression. \emph{The Annals of Statistics }%
\textbf{37} 35--72.

\item Fritsch, F.N.~\& Carlson, R.E. (1980). Monotone piecewise cubic
interpolation. \emph{SIAM Journal of Numerical Analysis }\textbf{17}
238--246.

\item Gervini, D. \& Gasser, T. (2004). Self-modeling warping functions. 
\emph{Journal of the Royal Statistical Society (Series B)} \textbf{66}
959--971.

\item Gervini, D. \& Gasser, T. (2005). Nonparametric maximum likelihood
estimation of the structural mean of a sample of curves. \emph{Biometrika} 
\textbf{92} 801--820.

\item Geyer, C.J.~(1994). On the asymptotics of constrained M-estimators. 
\emph{The Annals of Statistics }\textbf{22} 1993--2010.

\item Hall, P.~\& Horowitz, J. L. (2007). Methodology and convergence rates
for functional linear regression. \emph{The Annals of Statistics} \textbf{35}
70--91.

\item James, G.M. (2007). Curve alignment by moments. \emph{The Annals of
Applied Statistics} \textbf{1} 480--501.

\item James, G., Hastie, T. G. \& Sugar, C. A. (2000). Principal component
models for sparse functional data.\ \emph{Biometrika} \textbf{87} 587--602.

\item James, G., Wang, J. \& Zhu, J. (2009). Functional linear regression
that's interpretable. \emph{The Annals of Statistics} \textbf{37} 2083--2108.

\item Jupp, D. L. B. (1978). Approximation to data by splines with free
knots. \emph{SIAM J. Numer. Anal.} \textbf{15} 328--343.

\item Kneip, A., Li, X., MacGibbon, B. \& Ramsay, J.O. (2000). Curve
registration by local regression. \emph{Canadian Journal of Statistics} 
\textbf{28} 19--30.

\item Kneip, A. \& Ramsay, J.O. (2008). Combining registration and fitting
for functional models. \emph{Journal of the American Statistical Association}
\textbf{103} 1155--1165.

\item Liu, X. \& M\"{u}ller, H.-G. (2004). Functional convex averaging and
synchronization for time-warped random curves. \emph{Journal of the American
Statistical Association} \textbf{99} 687--699.

\item Magnus, J.R.~\& Neudecker, H. (1999). \emph{Matrix Differential
Calculus with Applications in Statistics and Econometrics (Second Edition)}.
New York: Wiley.

\item M\"{u}ller, H.-G., Chiou, J.-M., \& Leng, X. (2008). Inferring gene
expression dynamics via functional regression analysis. \emph{BMC
Bioinformatics} \textbf{9} 60.

\item Pollard, D. (1984). \emph{Convergence of Stochastic Processes.}
Springer, New York.

\item Pollard, D. (1990). \emph{Empirical Processes: Theory and Applications}%
. Hayward, California: Institute of Mathematical Statistics.

\item Ramsay, J.O. (1988). Monotone regression splines in action (with
discussion). \emph{Statistical Science }\textbf{3} 425--461.

\item Ramsay, J.O. \& Li, X. (1998). Curve registration. \emph{Journal of
the Royal Statistical Society (Series B)} \textbf{60} 351--363.

\item Ramsay, J.O. \& Silverman, B. (2005). \emph{Functional Data Analysis
(Second Edition).} Springer, New York.

\item Rockafellar, R.~\& Wets, R. (1998). \emph{Variational Analysis}. New
York: Springer.

\item Tang, R.~\& M\"{u}ller, H.-G. (2008). Pairwise curve synchronization
for functional data. \emph{Biometrika} \textbf{95} 875--889.

\item Tang, R.~\& M\"{u}ller, H.-G. (2009). Time-synchronized clustering of
gene expression trajectories. \emph{Biostatistics} \textbf{10} 32--45.

\item Yao, F., M\"{u}ller, H.-G. \& Wang, J.-L. (2005). Functional linear
regression analysis for longitudinal data. \emph{The Annals of Statistics }%
\textbf{33 }2873--2903.

\item Wang, K. \& Gasser, T. (1999). Synchronizing sample curves
nonparametrically. \emph{The Annals of Statistics} \textbf{27 }439--460.
\end{description}

\end{document}